\documentclass[letterpaper,twocolumn,10pt]{article}

\usepackage{usenix}
\usepackage{graphicx}
\usepackage{cite}
\usepackage{xcolor}
\usepackage{xifthen}
\usepackage{hyperref}
\usepackage{listings}
\usepackage{xcolor}
\usepackage{multirow}
\usepackage{tikz}
\usepackage{array}
\usepackage{subfig}
\usepackage{amssymb}
\usepackage{longtable}
\usepackage{soul}
\usepackage{xspace}

\usetikzlibrary{arrows,automata}

\date{}

\title{\Large\bf Characterizing Physical Memory Fragmentation}

\author{
{\rm Mark Mansi}\\
University of Wisconsin-Madison
\and
{\rm Michael M. Swift}\\
University of Wisconsin-Madison
}

\newcounter{observationn}
\newcommand{\finding}[1]{\smallskip\refstepcounter{observationn}\par\noindent\fbox{\parbox{0.97\linewidth}{\noindent\textbf{Finding~\theobservationn.} {\em #1}}}\smallskip}

\newcommand{\codesm}[1]{{\small \texttt{#1}}}

\newcommand{\sys}{{\texttt{andúril}\xspace}}
\newcommand{\syssumf}{{\texttt{andúril}\xspace}}

\newif\ifcomments
\commentsfalse

\ifcomments
  \newcommand{\todo}[1][]{ \ifthenelse{\isempty{#1}}{{\color{red}[TODO]}}{{\color{red}[TODO: #1]}} }
  \newcommand{\mike}[1]{ {\color{red}[MIKE: #1]} }
  \newcommand{\mnote}[1]{ {\color{red}[MIKE: #1]} }
  \newcommand{\markm}[1]{ {\color{green}[MARK: #1]} }
  \newcommand{\sujay}[1]{ {\color{brown}[SUJAY: #1]} }
  \newcommand{\jing}[1]{ {\color{brown}[JING: #1]} }
  \newcommand{\anthony}[1]{ {\color{brown}[ANTHONY: #1]} }
  \newcommand{\tomark}[1]{ {\color{orange}[To MARK: #1]} }
  \newcommand{\tomike}[1]{ {\color{blue}[To MIKE: #1]} }
  \newcommand{\reviewer}[1]{ {\color{purple}[REVIEWER: #1]} }
\else
  \newcommand{\todo}[1][]{}
  \newcommand{\tomark}[1]{}
  \newcommand{\tomike}[1]{}
  \newcommand{\sujay}[1]{}
  \newcommand{\jing}[1]{}
  \newcommand{\anthony}[1]{}
  \newcommand{\mike}[1]{}
  \newcommand{\mnote}[1]{}
  \newcommand{\markm}[1]{}
  \newcommand{\reviewer}[1]{}
\fi

\newcommand{\mypar}[1]{\smallskip\par\noindent\textbf{#1}~~}

\begin{document}

\maketitle

\begin{abstract}

External fragmentation of physical memory occurs when adjacent differently sized regions of allocated physical memory are freed at different times, causing free memory to be physically discontiguous.
It can significantly degrade system performance and efficiency, such as reducing the ability to use huge pages, a critical optimization on modern large-memory system.
For decades system developers have sought to avoid and mitigate fragmentation, but few prior studies quantify and characterize it in production settings.

Moreover, prior work often artificially fragments physical memory to create more realistic performance evaluations, but their fragmentation methodologies are \textit{ad hoc} and unvalidated.
Out of 13 papers, we found 11 different methodologies, some of which were subsequently found inadequate.
The importance of addressing fragmentation necessitates a validated and principled methodology.

Our work fills these gaps in knowledge and methodology.
We conduct a study of memory fragmentation in production by observing 248 machines in the Computer Sciences Department at University of Wisconsin - Madison for a week.
We identify six key memory usage patterns, and find that Linux's file cache and page reclamation systems are major contributors to fragmentation because they often obliviously break up contiguous memory.
Finally, we create \sys{}, a tool to artificially fragment memory during experimental research evaluations.
While \sys{} ultimately fails as a scientific tool, we discuss its design ideas, merits, and failings in hope that they may inspire future research.
\end{abstract}

\section{Introduction and Motivation}

In the era of cloud computing, long-running and resource-hungry services, and increasing memory capacities, memory fragmentation plays an increasing role in system performance.
Fragmentation causes memory allocations to fail unnecessarily, which can cause system-wide inefficiencies (e.g., needing to over-provision machines) or hinder the operation of crucial optimizations (e.g., huge pages~\cite{panwarMakingHugePages2018}).
For example, fragmentation degrades application performance on Linux by 50\% to 500\% due to failed or delayed attempts to create huge pages~\cite{mansiCBMMFinancialAdvice2022}.
Moreover, numerous hardware proposals aim to make use of ``incidental contiguity'' -- unintentional physical contiguity resulting from kernel implementation, such as the use of buddy allocators~\cite{phamCoLTCoalescedLargeReach2012,parkHybridTLBCoalescing2017}.

In particular, we focus on \textit{external} fragmentation of \textit{physical} memory managed by the OS kernel.
\textit{External memory fragmentation} is a system state in which free memory is too discontiguous to satisfy allocations, even though the total amount of free memory is larger than the requested amount.
\jing{Are these well-known definitions? Perhaps cite.}
(It is distinct from \textit{internal memory fragmentation}, where an allocated block of memory is underutilized).
The diversity of allocation sizes and the increasing reliance on large contiguous physical memory allocations exacerbate external memory fragmentation.

However, surprisingly little prior work exists to characterize or quantify fragmentation, even though many papers refer to its impact and many designs seek to mitigate or avoid it.
For example, it is unclear what contiguous memory allocations are actually used for, the degree to which workloads fragment system memory, what types of memory allocations cause fragmentation, how fragmentation changes over time, etc.

\begin{table*}
\footnotesize
\centering
\begin{tabular}{|p{2in}|p{2in}|p{2in}|}
\hline
\textbf{Methodology} & \textbf{Pros} & \textbf{Cons} \\ \hline
Do nothing/run on a freshly rebooted machine [most work in architecture]
    & Easy to understand and reproduce
    & Not realistic
    \\ \hline
Run experiments on office workstations that have not been rebooted in 2 months~\cite{phamCoLTCoalescedLargeReach2012}
    & Realistic
    & Takes 2 months; not well-defined or reproducible \\ \hline
Run all workloads without rebooting between~\cite{gormanSupportingSuperpageAllocation2008,gormanWhatWhyWhere2006}
    & Maybe realistic
    & Not well-defined or reproducible; level of fragmentation changes between workloads \\ \hline
Run each workload multiple times before recording results; increase the \codesm{MAX\_ORDER} of the allocator~\cite{yanTranslationRangerOperating2019}
    & Well-defined
    & Unlikely to be realistic; not clear if rerunning a workload increases fragmentation \\ \hline
Run userspace workload that allocates all memory and free parts of it~\cite{kwonCoordinatedEfficientHuge2016,michailidisMEGAOvercomingTraditional2019}
    & Well-defined; easy to understand
    & Unrealistic: likely produces pathological fragmentation; does not produce kernel-level unmovable pages \\ \hline
Run \codesm{memhog}~\cite{NumactlNumactl2021} (similar to above, but touches allocated memory randomly)~\cite{bhattacharjeeTranslationTriggeredPrefetching2017,coxEfficientAddressTranslation2017,alvertiEnhancingExploitingContiguity2020}
    & Well-defined; easy to understand; potentially less pathological
    & Maybe unrealistic; does not produce kernel-level unmovable pages \\ \hline
Run \codesm{memhog}~\cite{NumactlNumactl2021} after shuffling page lists~\cite{yanTranslationRangerOperating2019}
    & Well-defined; easy to understand; more likely to introduce persistent fragmentation
    & Maybe unrealistic;  does not produce kernel-level unmovable pages \\ \hline
Read “several” files then start experiment while memory still allocated~\cite{panwarHawkEyeEfficientFinegrained2019}
    & Induces kernel-level unmovable pages; easy to understand
    & Ignores anonymous memory; underspecified in paper and artifact \\ \hline
Repeatedly allocate file-backed and anonymous \codesm{mmap}s~\cite{panwarMakingHugePages2018}
    & Considers mutliple sources/types of fragmentation
    & Underspecified in paper and artifact \\ \hline
Allocate until memory pressure; partially free; repeat only with the target amount of memory~\cite{zhuComprehensiveAnalysisSuperpage2020}
    & Well-defined; well-specified in paper; creates persistent fragmentation
    & does not produce kernel-level unmovable pages \\ \hline
Simulation with memory usage “profile” generated by RNG~\cite{parkHybridTLBCoalescing2017}
    & Well-defined; easy to understand; reproducible
    & Only works for simulation; unclear how realistic \\ \hline
\end{tabular}
\caption{Fragmentation methodologies found in prior work.\label{fig:priorfrag}}
\end{table*}

Additionally, because fragmentation significantly impacts system behavior, experimentally reproducing it is important.
Prior work in kernel-level memory management artificially fragments memory using many different methodologies to make experiments more realistic, as shown in Table \ref{fig:priorfrag}.
Unfortunately, their methodologies are \textit{ad hoc}, unvalidated, and inconsistent.
We found many of them difficult to reproduce, with results that vary wildly from run to run, making scientific interpretation of results difficult.

To fill the above gaps in knowledge, we conduct the first study of external physical memory fragmentation in productions settings:

\vspace{-0.2cm}
\begin{itemize}
\itemsep-0.1cm
\item We instrument 248 live systems in the CS department at the University of Wisconsin - Madison, including departmental infrastructure and nodes from the Center for High-Throughput Computing.
\item We analyze the amount, usage, and contiguity of free and allocated physical memory in the system over 1 week.
\item We find that Linux uses contiguous memory mostly for huge pages (512 contiguous pages), buffers, and kernel slabs for data structures (2-8 contiguous pages).
\item We identify several common memory usage patterns.
\item We identify the file cache and reclamation algorithm as key sources of fragmentation.
\item We quantify challenges and opportunities for huge page usage.
\item We analyze memory usage over time and suggest ways to optimize fragmentation control.
\item To improve the state of experimentation, we set out to create a new tool, \sys{}, that artificially and configurably fragments physical memory.
\item While \sys{} ultimately fails as a scientific tool, we discuss its design ideas, merits, and failings in hope that they may inspire future research.
\end{itemize}

\section{Related Work}

\textit{Memory fragmentation} is a system state where memory allocations fail even though the total amount of unused memory exceeds the allocations' request sizes.
\textit{External} fragmentation occurs when free memory is too discontiguous to satisfy requests: when adjacent differently sized allocations have differing lifetimes, free memory will interleave with allocated memory.
Both userspace and kernel memory managers suffer framgentation, but we focus on \textit{kernel} memory management for \textit{physical memory}.

Prior work extensively studies usermode allocators, including internal and external fragmentation control, use of huge pages, and more~\cite{hunterMallocEfficiencyFleet2021,evansScalableMemoryAllocation2011,powersMeshCompactingMemory2019,robsonWorstCaseFragmentation1977,johnstoneMemoryFragmentationProblem1998}.
Additionally, recent work has examined huge page usage and proposed mechanisms for tuning existing defragmentation mechanisms in Linux or BSD~\cite{zhuComprehensiveAnalysisSuperpage2020,kwonCoordinatedEfficientHuge2016,panwarHawkEyeEfficientFinegrained2019,panwarMakingHugePages2018}.
Our work instead focuses on characterizing and reproducing physical memory external fragmentation itself, and can be used to inform and help evaluate such work.
Surprisingly little prior work exists on this topic, despite the fact that many papers acknowledge its impact; we believe this is because large contiguous physical memory allocations have been rare in practice until recently.

\mypar{Artificial Fragmentation.}
In order to study the performance of new memory management proposals in a realistic setting, researchers often artificially induce fragmentation.
Table \ref{fig:priorfrag} shows 11 different methodologies we found in prior work.
As the table shows, the approaches are diverse.
Some prior work qualitatively describes the benefits of their methodology compared to prior work (e.g., \cite{panwarMakingHugePages2018,zhuComprehensiveAnalysisSuperpage2020}).
Unfortunately, none of these methodologies are validated, and many are difficult to reproduce, making results scientifically incomparable and difficult to understand.
No prior work analyzes or validates the state of the system after after artificial fragmentation.
\sys{} is our attempt to fill the need for a validated, easy-to-use artificial physical memory fragmentation methodology.
We hope that our exploration can help lay the groundwork for more scientific research in the future.

\mypar{Fragmentation control mechanisms} operate either at allocation time (as part of the allocator) or via reclamation/migration (independently from the allocator).
Linux and FreeBSD use \textit{buddy allocators}, which manage memory in power-of-two-aligned-and-sized blocks of base pages.
The alignment and size constraints enable identifying and merging contiguous free memory blocks without significant metadata.
Merging free blocks naturally preserves contiguity by segregating memory free lists by contiguity; smaller allocations are satisfied from less contiguous regions, leaving larger regions intact longer~\cite{phamCoLTCoalescedLargeReach2012}.
FreeBSD features a \textit{reservation} system that defers the allocation of potential huge pages as long as possible~\cite{navarroPracticalTransparentOperating2002}.
Linux features a ``compaction'' (defragmentation) daemon that migrates movable memory contents away from otherwise contiguous memory.

\mypar{Software Aging.}
Prior work has studied \textit{software aging} -- the gradual degradation of software behavior due to the accrual of error during execution.
Memory fragmentation is a form of software aging, but most prior work examines buggy behavior such as resource leaks (e.g., memory, file descriptors, locks), rather than fragmentation~\cite{araujoExperimentalEvaluationSoftware2011, cotroneoSurveySoftwareAging2014,matiasMeasuringSoftwareAging2010}.

The storage community has examined \textit{file system aging} -- the degradation of the file system's physical layout on disk over time.
Unlike memory fragmentation, file system aging leads to poor performance when \textit{accessing} data due to the physical organization of storage devices (e.g., more HDD seeks or poor usage SSD internal parallelism).
Prior work focuses on accurately and portably recreating an aged disk layout for experimental purposes~\cite{smithFileSystemAging1997, conwayFilesystemAgingIt2019}, usually by synthesizing/recording traces of high-level operations (e.g., file system calls) and replaying them~\cite{conwayFilesystemAgingIt2019,conwayFilesystemAgingIt2019,joukovAccurateEfficientReplaying2005,agrawalGeneratingRealisticImpressions2009,conwayFileSystemsFated2017,weissROOTReplayingMultithreaded2013,zhuTBBTScalableAccurate2005}.
File system aging requires significant time, space, and effort to replay traces.
In contrast, \sys{} aims to trade off some accuracy in exchange for being fast to deploy and use, and not requiring significant auxiliary data or processing.

\begin{table*}[t]
\footnotesize
\centering
\begin{tabular}{|p{1in}|p{2.0in}|p{3.0in}|}
\hline
\textbf{ Process}            & \textbf{ Description}                    & \textbf{ Contiguous Memory Usage (Size per Allocation in Pages)} \\ \hline
\codesm{java}      & Java virtual machine           & THP (512), Slab(memory mapping, process) (2-8) \\ \hline
\codesm{jbd}       & ext4 journaling kernel thread  & Slab(block I/O) (2-8) \\ \hline
\codesm{lettuce-epol} & Asynchronous I/O (\codesm{redis}) & UDS (2-4), THP (512) \\ \hline
\codesm{mcf\_s}    & SPEC 2017 mcf                  & THP (512) \\ \hline
\codesm{memcached} & In-memory key-value store      & THP (512) \\ \hline
\codesm{pgrep}     & Searches for processes by name  & Slab(file,dirent,path) (2-8) \\ \hline
\codesm{redis}     & Redis key-value store server   & UDS (2-4), THP (512), Slab(filesystem) (8) \\ \hline
\codesm{sshd}      & Secure Shell daemon (server)   & Networking packetization buffer (8), Slab(TCP, memory mapping) (2-8) \\ \hline
\codesm{sudo}      & Execute as superuser           & Slab(memory mapping, file, path) (2-8), THP (512) \\ \hline
\codesm{xz\_s}     & SPEC 2017 xz                   & THP (512), Slab(file,dirent,path) (2-8) \\ \hline
\end{tabular}
\caption[Contiguous memory usage of notable processes in various workloads]
{Contiguous memory usage (in order of number of allocations) of notable processes in workloads. THP = Transparent Huge Pages. UDS = Unix Domain Socket buffer. Slab(X) = kernel slab allocator for X data structures.\label{tab:hallocapp}}
\end{table*}

\section{Study: Linux Allocation Sizes \label{sec:sizestudy}}

It is unknown what allocation sizes are common, and thus what causes and is impacted by fragmentation.
We use BPF to record all contiguous multi-page allocations in Linux on server-class machines using the same experimental setup to Mansi \textit{et al}~\cite{mansiCBMMFinancialAdvice2022} for the workloads \codesm{memcached}, \codesm{xz}, \codesm{mcf}, and \codesm{dc-mix} -- a variety of server and batch workloads.

\finding{In Linux, the most common use of contiguous memory is huge pages (512 contiguous pages).
I/O-heavy processes also use contiguous memory for buffers and slab allocators for kernel data structures (usually 2-8 contiguous pages).
(Table \ref{tab:hallocapp})
}

Typically, a single process's use of contiguity was dominated by one use case.
In \codesm{memcached}, \codesm{mcf}, and \codesm{xz}, huge pages comprise 84\%, 48\%, and 86\% of all large allocations.
In \codesm{redis}, almost 100\% of large allocations back a Unix Domain Socket buffer for threads to communicate.
In \codesm{sshd}, 87\% of contiguous memory is for network data packetization buffers.
All processes triggered kernel slab allocations for kernel data structures, which have been shown to increase fragmentation~\cite{panwarMakingHugePages2018}.
Daemons, such as ext4's journaling thread, also make slab allocations (e.g., I/O buffers).

\textit{ Implications: } Though Linux maintains intermediate-sized free lists, they are rarely used except to fill lower-order lists.
Separating the allocation mechanisms for 2-8 page chunks and huge pages may preserve contiguity for longer.
For example, smaller allocations could be made from a different part  of the physical address space from huge pages.

\section{Study: Real World Fragmentation \label{sec:fragstudy}}

In order to gain a better understanding of fragmentation in the real
world, we instrument live machines in the Computer Sciences department at University of Wisconsin-Madison, including departmental infrastructure and infrastructure and compute nodes from the Center for High-Throughput Computing.

\subsection{Methodology}

We observed 248 instrumented machines total, from department infrastructure (DI) and high-performance computing (HPC) infrastructure, spanning several generations and manufacturers of x86 hardware (Intel Blackford '05 and AMD Opteron '09 to Intel Cascade Lake '20 and AMD EPYC'19), and different sizes of memory (from 1GB to 4TB).
We observed 45 different classes of workloads (as classified by system administrators; e.g., file servers, web servers, hypervisors, research workstations, and compute nodes from a high-performance computing cluster), detailed in Table \ref{fig:nodes} (in the Appendix).
Many DI nodes are virtual machines, whereas most HPC nodes are bare-metal hardware.
DI nodes mostly run Ubuntu 20.04 with kernel 5.4; HPC nodes run Centos7 with kernel 3.10.0 or 5.0.8.

We instrumented all systems for 7 days from April 6 to April 13, 2022.
In every 30-minute window during this period, we pick a random minute to perform a ``snapshot'' of the system\footnote{Selecting a random minute avoids pathologically missing events that happen at a regular interval.}.
Our instrumentation captures the contents of \codesm{/proc/kpageflags}, which contains information about each physical 4KB page of memory and is our primary view into the memory fragmentation of the system.
We also capture the kernel version, hardware details, and uptime.
For VMs, we capture all information at the guest OS level.
No machines were captured at both the guest and host OS level.

\subsection{Free Memory \label{sec:free}}

We examine the amount of free memory on all machines to better understand the impact of fragmentation.

\finding{Small VMs experience high memory pressure and churn.
Machines with large amounts of memory often have lower fragmentation.
}

DI nodes tend to use small VMs ($\le 4$GB total) with $\le 20\%$ free memory.
Most DI workloads and several HPC workloads have less than 2GB of free memory at least 75\% of the time.
These small VMs are sized to have minimal idle memory (e.g., by using balloon drivers).
Background tasks, such as logging or virus scanning, have high file cache usage that leads to churn (i.e., pages being reclaimed and recycled) even if load is low.

Machines with more memory are more likely to be either highly idle (i.e., most cores idle; $\ge 50\%$ free memory) or highly utilized ($\le 10\%$ free memory).
Often large amounts of memory are freed after a process terminates; as we will see later, this leads to lower fragmentation.

\textit{Implications:}
On small VMs, kernels may want to use more passive fragmentation control, such as careful page (re)placement, because memory pressure is high and resources are limited.
On large machines with more resources, more aggressive fragmentation control (e.g., compaction) may be acceptable.

\begin{figure*}[t!]
\centering
\includegraphics[width=0.65\textwidth]{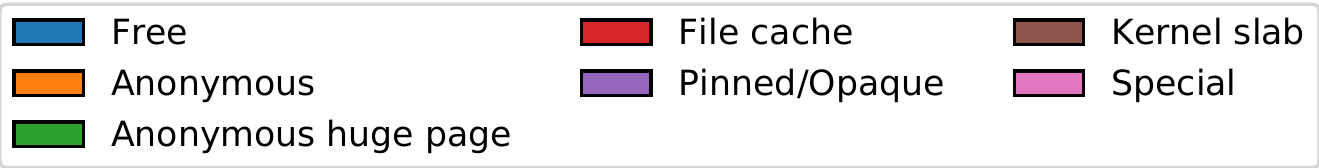} \\
\subfloat[Slow-filling buffers]{\includegraphics[width=0.31\textwidth]{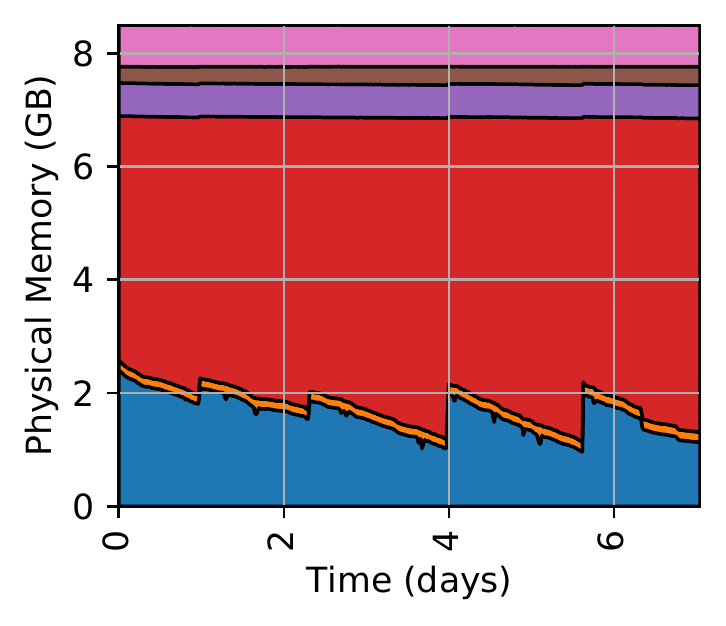}\label{fig:exmem-mars02}}
\subfloat[GPU workload]{\includegraphics[width=0.33\textwidth]{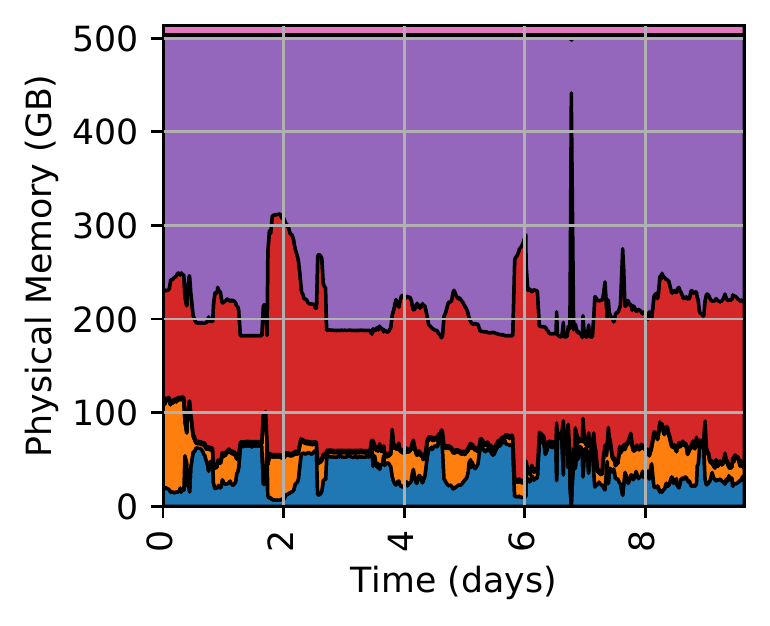}\label{fig:exmem-gpu2001}}
\subfloat[Small VM]{\includegraphics[width=0.33\textwidth]{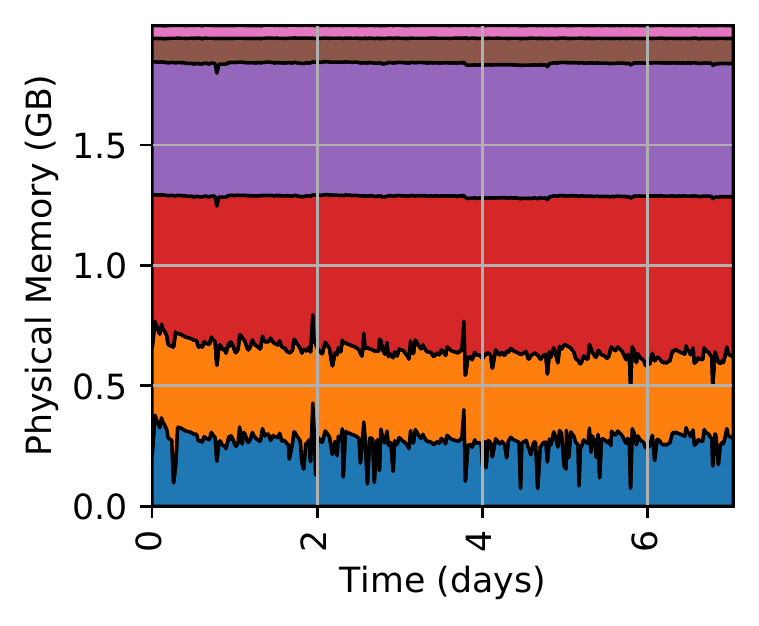}\label{fig:exmem-dns2}} \\
\subfloat[Highly variable]{\includegraphics[width=0.33\textwidth]{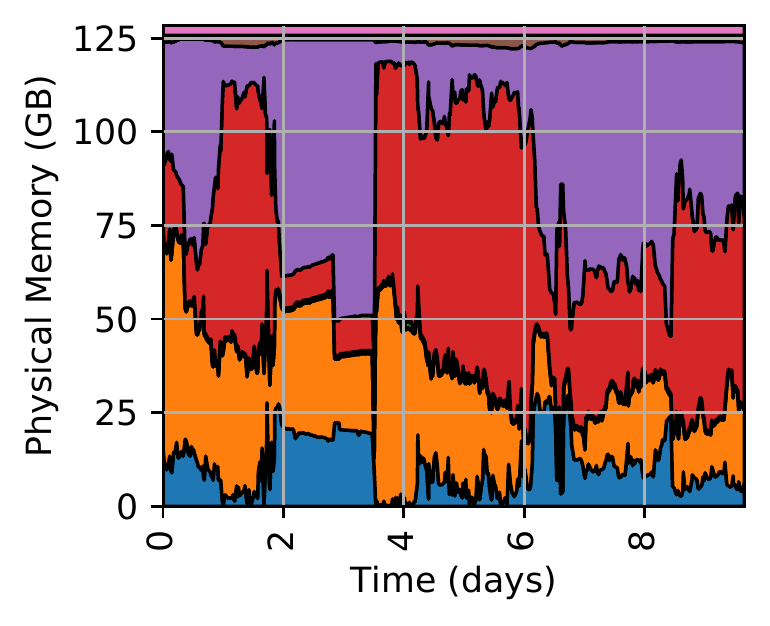}\label{fig:exmem-e1017}}
\subfloat[Idle]{\includegraphics[width=0.32\textwidth]{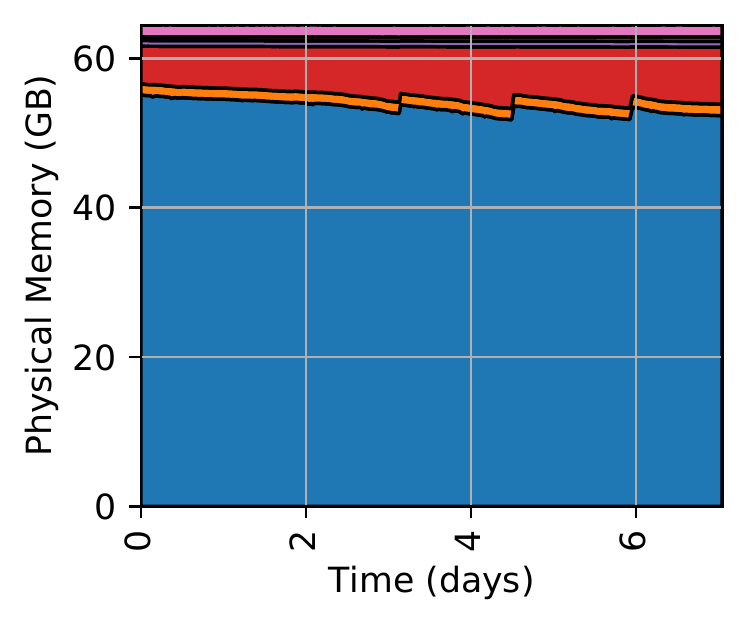}\label{fig:exmem-durga}}
\subfloat[Hypervisor]{\includegraphics[width=0.32\textwidth]{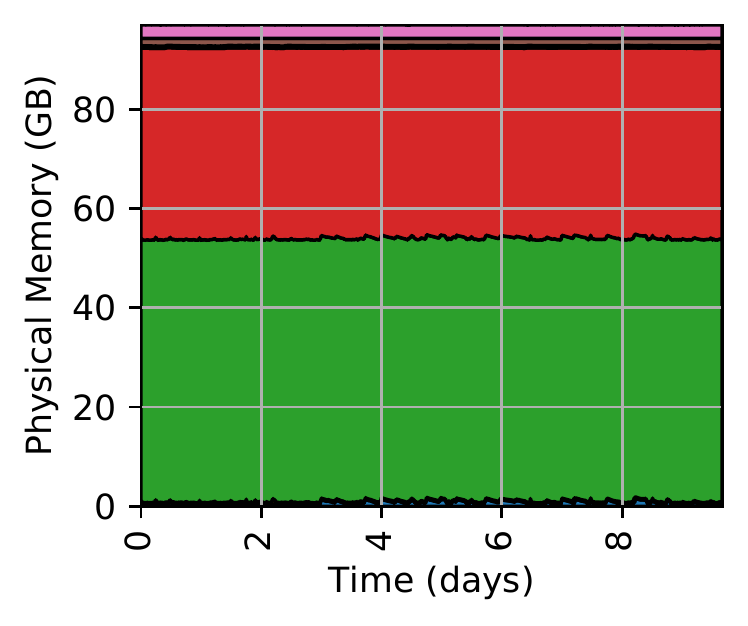}\label{fig:exmem-h3000}}
\caption{Examples of memory usage patterns. \label{fig:exmemusage}}
\end{figure*}

\begin{figure*}[t!]
\centering
\includegraphics[trim={1cm, 1cm, 4.8cm, 1cm},  height=5cm, angle=90]{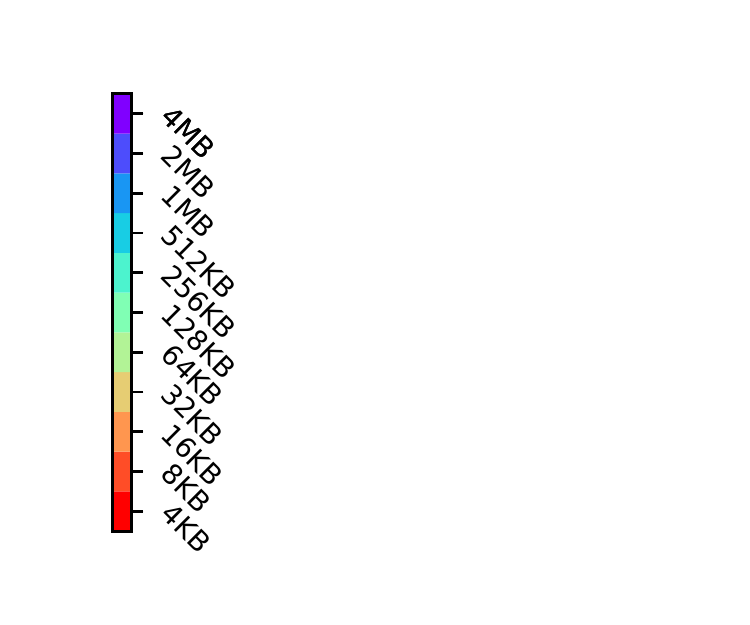}  \\
\subfloat[Slow-filling buffers]{\includegraphics[width=0.31\textwidth]{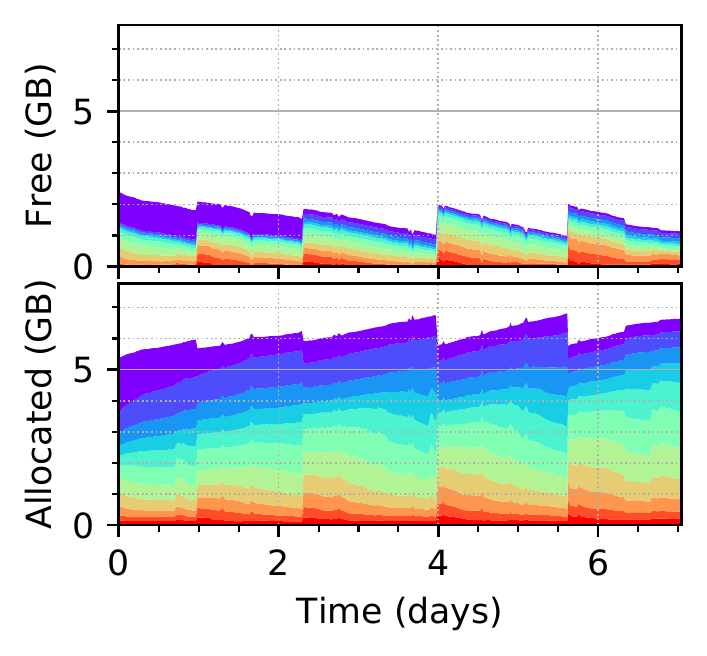}\label{fig:exdist-mars02}}
\subfloat[GPU workload]{\includegraphics[width=0.33\textwidth]{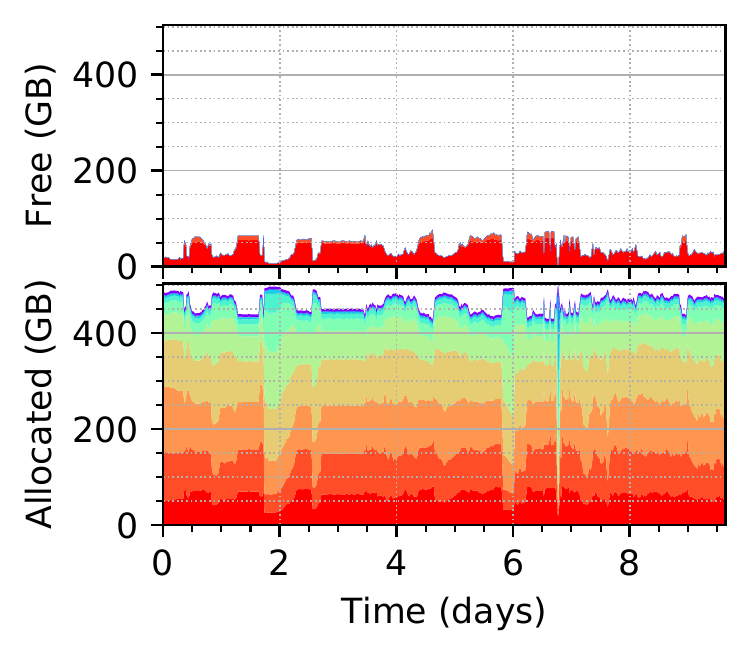}\label{fig:exdist-gpu2001}}
\subfloat[Small VM]{\includegraphics[width=0.33\textwidth]{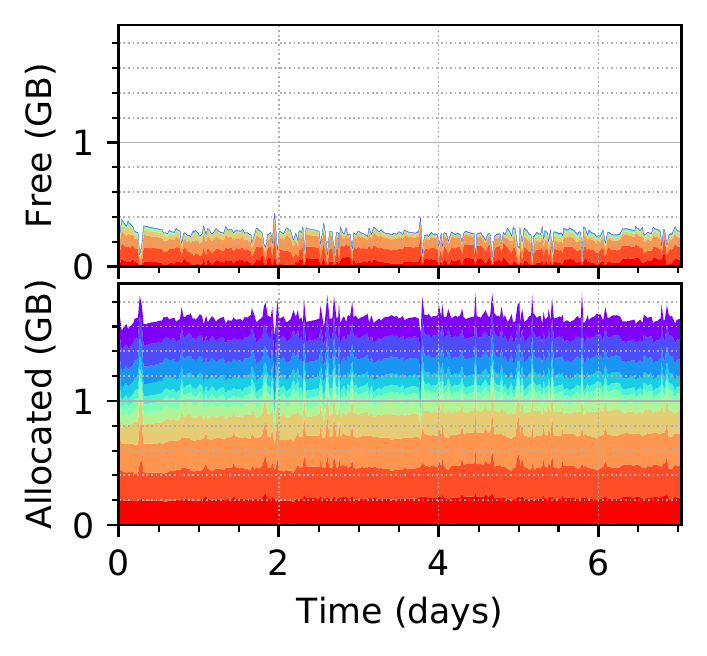}\label{fig:exdist-dns2}} \\
\subfloat[Highly variable]{\includegraphics[width=0.33\textwidth]{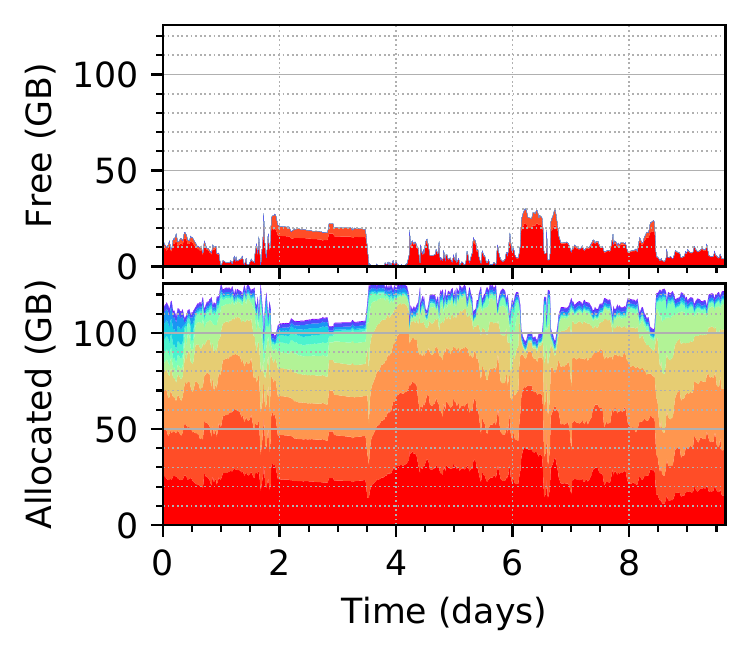}\label{fig:exdist-e1017}}
\subfloat[Idle]{\includegraphics[width=0.32\textwidth]{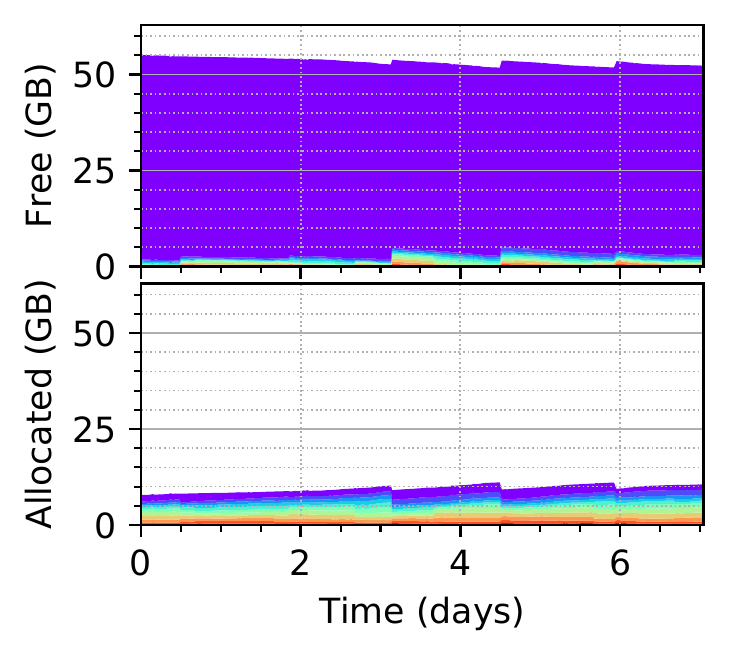}\label{fig:exdist-durga}}
\subfloat[Hypervisor]{\includegraphics[width=0.32\textwidth]{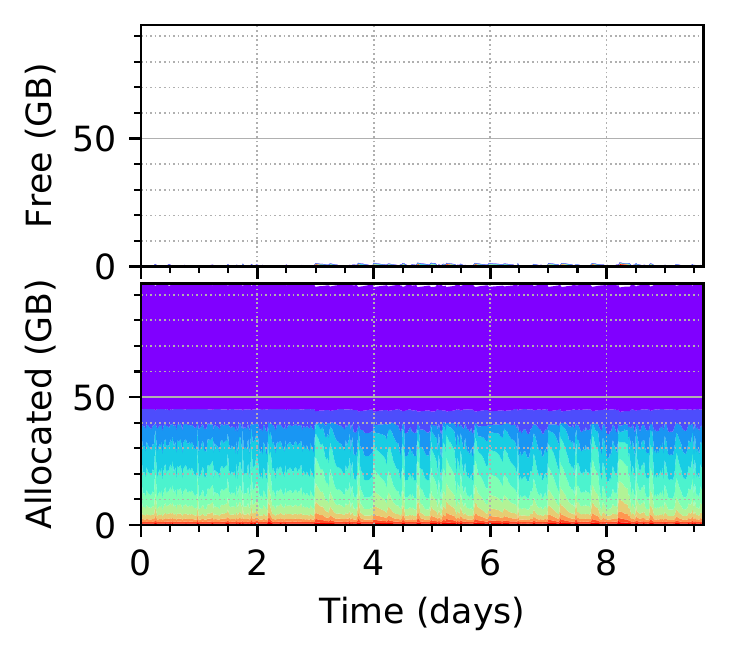}\label{fig:exdist-h3000}}
\caption[Amount of memory contiguity for different memory usage patterns]
{Amount of contiguity in free and allocated memory for the different memory usage patterns in Figure \ref{fig:exmemusage}.\label{fig:exfragdist}}
\end{figure*}

\subsection{Memory Usage and Homogeneity}

We characterize machines' use of physical memory, and analyze the contiguity of physical memory with the same page flags in \codesm{/proc/kpageflags} -- which we call ``homogeneous''.
Homogeneity is a good proxy for contiguity/fragmentation because contiguous memory allocations must be homogeneous in usage and page table permissions; non-homogeneous regions indicate discontiguous/fragmented memory usage, so homogeneity bounds contiguity.

\finding{We observed 6 common and distinct memory usage patterns (Figure \ref{fig:exmemusage}).}

In Linux, we found that all memory in the system can be classified as one of seven usage types: free/unallocated memory, anonymous (non-huge-page) memory, anonymous huge page memory, file cache memory, pinned/opaque memory (e.g., used by drivers/DMA), kernel slab memory (used for common kernel data structures), and other special cases (zero pages, pages with detected hardware failures, etc).

We observed 6 common memory usage patterns:

\vspace{-0.3cm}
\begin{enumerate}
\itemsep-0.1cm
\item {\em Slow-filling buffers}: slow-filling buffers that periodically (usually 1-2 times daily) flush to disk. 
\item {\em GPU workloads}: dozens of gigabytes of memory dedicated to I/O. 
\item {\em Small VMs with low-medium load}: small single-task dedicated VMs with fairly consistent memory usage, only a few hundred MBs of free memory, and significant memory dedicated to balloon drivers~\cite{esx}. 
\item {\em Highly variable and unpredictable workloads}: widely varying and unpredictable memory usage as compute tasks run, terminate, and are replaced. 
\item {\em Idle}: large amounts of idle free memory; low activity. 
\item {\em Hypervisors}: mostly huge pages backing guest memory.
\end{enumerate}

\textit{Implication:} Memory usage is highly diverse but often predictable.
While current designs assume workloads are unknown, often this assumption is overly conservative.

\finding{Memory reclamation, triggered by memory pressure, is a major source of fragmentation because it breaks up homogeneous regions.
The fragmentation is often irreparable, even when 1/4 of memory is freed periodically.
}

This is a key finding of our work.
Memory pressure causes fragmentation to irreparably worsen by up to an order of magnitude, even if up to 1/4 of total memory is periodically freed, supporting prior findings~\cite{0sim}.
Often such freed pages are quickly allocated to the file cache, preventing them from being merged.
Batch tasks and background daemons/loggers can cause regular and periodic memory pressure, leading to increasing fragmentation over time.
These behaviors are especially clear in patterns (1) - (4), as exemplified in Figures \ref{fig:exdist-mars02} to \ref{fig:exdist-e1017}.
Unfortunately, as shown in Section \ref{sec:free}, memory pressure is common.

Conversely, homogeneity (of both allocated and free memory) is correlated to free memory ($R^2=0.63$, i.e., free memory is the single most important factor in amount of homogeneity).
Machines that are never fully utilized tend to retain significant amounts of homogeneity in both free and allocated memory.
Idle machines tend not become significantly more or less fragmented over time; there is no memory pressure to trigger Linux's reclamation (increasing fragmentation) or compaction (decreasing fragmentation).
Also, most idle machines lack significant fragmentation in their free memory anyway.

\textit{Implication:} Reclamation must be made contiguity-aware, so that it does not needlessly break up memory regions.

\finding{
File caches comprise significant memory usage on most machines.
Memory allocated to file caches usually becomes progressively more fragmented over time, especially in the presence of reclamation.
\label{fnd:file}}

This is a key finding of our work, illustrated by corresponding Figures \ref{fig:exmemusage} and \ref{fig:exfragdist}.
In patterns (1), (2), and (4), significant file cache activity leads to reclamation, which as we show later, increases fragmentation (illustrated by corresponding Figure \ref{fig:exfragdist}).
Moreover, Linux lacks general huge page support in the file cache; the file cache manages memory at base page granularity, so memory management operations often obliviously break up ``incidentally contiguous''~\cite{phamCoLTCoalescedLargeReach2012} memory regions.
All patterns except (5) use significant amounts of file cache memory, so file cache memory has a significant role in fragmentation control.

\textit{Implications:} Kernels should make file caches contiguity-aware.
File cache huge page support would reduce fragmentation and allow this significant memory pool to use huge pages.
Interestingly, most prior work focuses on anonymous huge pages~\cite{navarroPracticalTransparentOperating2002,gormanSupportingSuperpageAllocation2008,kwonCoordinatedEfficientHuge2016,panwarHawkEyeEfficientFinegrained2019,panwarMakingHugePages2018,zhuComprehensiveAnalysisSuperpage2020,yanTranslationRangerOperating2019}.
Additionally, pages that reduce fragmentation can be prioritized for eviction from the file cache, including during writeback operations.
For example, the kernel can flush dirty pages that interrupt a run of otherwise clean pages, or reclaim pages that are already not physically contiguous with their neighbors so as to avoid worsening fragmentation.

\finding{
Kernel slab memory and page tables usually accounted for very little memory usage.
}

Combined, slab and page table memory comprised $\le5\%$ on 44\% of machines and $\le 13\%$ on all remaining machines other than 6 outliers.

\textit{Implication:} While some prior work significantly modifies the memory allocator to avoid fragmentation by these pages~\cite{panwarMakingHugePages2018}, it may suffice (and be more efficient) to reserve a small dedicated memory pool for slabs and page tables, falling back to the normal allocator in case of higher usage.

\finding{For each finding above, there was one or two exceptional nodes that broke the rule.}

For example, there was one node that had high utilization and but also relatively high homogeneity.
There was also a small but significant minority of nodes that did not fit any of the six patterns we identified, and a handful of nodes that had sizable amount of kernel slab memory.

\textit{Implications:} It may make sense to have specialized memory management policies for common behaviors, but general fallback policies are needed for exceptions.

\begin{figure}[t!]
    \centering
    \includegraphics[width=\columnwidth]{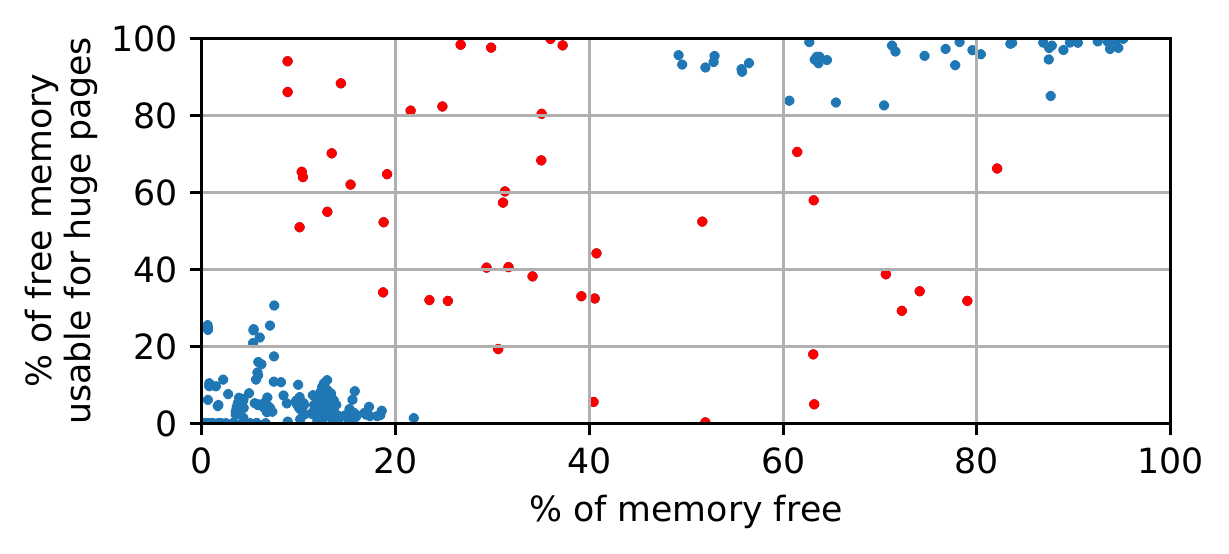}
    \caption[Median \% of free memory vs free memory contiguous enough for a 2MB huge page]
    {Median \% of free memory vs median \% of free memory that is contiguous enough for a 2MB huge page throughout the observation window for each machine.
    The 205 blue points indicate either low-memory-low-huge-page nodes or low-usage-high-huge-page nodes.
    The 43 red points indicate all other nodes. \label{fig:hp-feasibility}}
\end{figure}

\subsection{Huge Page Feasibility and Usage}

We examine huge page usage and the percentage of homogeneous regions that are 2MB or larger, a common huge page size on x86 and ARM.

\finding{Few machines used huge pages.}

Only 69 machines used more than 10MB of huge pages at any time throughout the observation window, and many of those machines did not use huge pages most of the time.
Only 35 machines used more than 1GB of huge pages at any time throughout the observation window.
Applications do not usually request huge pages explicitly, and most machines are configured not to use huge pages, surprisingly.
Except for hypervisor nodes, no machine had more than 1/4 of their memory backed by huge pages (and often much less).
Hypervisor nodes used huge pages for over 1/2 of memory, presumably to back guest pages.

\textit{Implication:} Kernels need transparent huge page systems that can be on-by-default without regressing performance.

\finding{Large-scale contiguity (e.g., huge pages) is rare, but smaller-scale contiguity is quite common.}

Figure \ref{fig:exfragdist} exemplifies this finding.
Small VMs usually have homogeneity up to 64KB, but rarely large than that.
High-variability machines, such as HPC compute nodes (e.g., Figure \ref{fig:exdist-e1017}), rarely had homogeneity greater than 128KB, especially in free memory.
Only when a task consuming more than 1/3 of memory terminated did fragmentation decrease, and then only 1/6 of memory became less fragmented.
Moreover, the unfragmented memory was often quickly consumed by Linux's Transparent Huge Page system, leaving the remaining free memory fragmented.
Kernel slab memory tends to have 8KB-16KB contiguity generally, which matches our findings in the previous section.

\textit{Implications:} Our findings support prior proposals for contiguity-aware TLBs~\cite{phamCoLTCoalescedLargeReach2012,parkHybridTLBCoalescing2017}.
They also suggest a 64KB granularity for some kernel mechanisms, as such homogeneous chunks are commonly present.
Such mechanisms might include memory ballooning/hot-unplugging, reclamation, range-based virtual-address translations or permissions, or even the kernel allocator.
Moreover, a smaller intermediate 64KB huge page size might be more usable than existing sizes (2MB, 1GB on x86\_64), which often require compaction.

\finding{On machines with little free memory, little of that free memory was usable for huge pages.
}

Figure \ref{fig:hp-feasibility} shows the amount of free memory on each machine that is contiguous enough to create a huge page compared to the total amount of free memory.
Over 83\% of machines fit in one of two general patterns:
(1) if $\le25\%$ of memory is free, most of it is discontiguous (lower left); or
(2) if $\ge50\%$ of memory is free, almost all of it is highly contiguous (upper right).

43 nodes (17\%; red in the figure) fit neither pattern.
Of these, most had a relatively large amount of file cache memory at some point.
Many were also machines with highly variable workloads.
A few also had relatively large amounts of slab memory (shown to cause fragmentation in the past~\cite{panwarMakingHugePages2018}) and huge pages (i.e., contiguous memory was already being used, leaving free memory fragmented).

The use of file cache pages on an otherwise idle machine can reduce contiguity, as discussed in Finding \ref{fnd:file}.
For example, about 15\% of large machines with ample free memory had low contiguity compared to similar machines.

\textit{Implication:} Currently in Linux, to have contiguity a machine must be either overprovisioned with memory or pay the overhead of memory compaction.

\subsection{Change over time}

We examine the usage of each page of memory on each machine across snapshots.

\finding{The page flags of $\ge 95\%$ of pages changed less than once every 3 hours, and on most machines, 20-60\% of pages did not change at all.}

On almost all machines, $\ge 95\%$ of pages changed usage (i.e., flags in \codesm{kpageflags})  $\le 50$ times over the course of a week (i.e., less than 1 change every 3 hours).
Moreover, on most machines, 20-60\% of pages did not change usage at all during the observation period.
While it is possible that additional changes were not captured between snapshots, we took our snapshots at a random time in each half hour window so as not to miss frequent or periodic behavior.

For pages that did change, the changes tended to be slightly bursty, rather than uniform over time -- on most machines, the distribution of time between changes was long-tailed (skewness between 2 and 4\footnotemark), indicating multiple relatively rapid changes followed by extended periods with no change.
\footnotetext{\textit{Skewness} is a statistical quantification of the long-tailed-ness of a distribution. 0 indicates a symmetric distribution, while positive and negative values indicate right- or left-tailed-ness, respectively. A common rule of thumb is that values $\ge 1$ or $\le -1$ indicate highly skewed distributions~\cite{bulmerPrinciplesStatistics1979}.}
Only 19 machines had a high rate of page changes.
These machines had little free memory and bursty workloads that drove significant file cache activity.

\textit{Implications:}
This finding suggests that fragmentation is disproportionately affected by a small set of rapidly-changing pages.
If so, then fragmentation control can be more efficient by focusing on these pages, rather than scanning through all pages, as current systems do.
However, further investigation is needed; because we captured snapshots at an average interval of 30 minutes, it is possible that our methodology misses important changes in between snapshots that affect fragmentation's evolution over time without changing overall page usage (e.g., the flushing of dirty pages in the file cache).

\subsection{Uptime}

\finding{
Uptime has little to do with fragmentation; rather memory pressure is a much more important factor than uptime in magnitude of memory fragmentation.
}

For security reasons, over half of DI machines restart weekly, while 91\% were restarted within two weeks.
In contrast, HPC machines had long and highly variable uptimes; over 97\% of machines had an uptime over one month, and nearly 20\% of machines had an uptime between 1 and 4 years.
Many DI machines have low contiguity despite short uptimes due to their small size and memory pressure.
Moreover, many of them had been rebooted just days before our observation window but were already as fragmented as long-uptime machines, despite very modest load and often entirely due to background processes such as logging or virus scans.
In contrast, many non-idle HPC machines had significantly longer uptimes but comparable or better memory contiguity.

\textit{Implications:}
While some prior work suggests rebooting as a fragmentation control measure~\cite{araujoExperimentalEvaluationSoftware2011}, our results show that this technique is unreliable at best.

\section{\syssumf{}}

Because fragmentation significantly impacts system behavior, experimentally reproducing it is important.
Unfortunately, as shown in Table \ref{fig:priorfrag}, prior methodologies are \textit{ad hoc}, unvalidated, and inconsistent.
They also use at least 3 different metrics for quantifying and describing fragmentation, and some use qualitivative descriptions only.

Broadly speaking, we observe three main approaches to artificial fragmentation.
The first and simplest is to do nothing: always run on a freshly rebooted, unfragmented machine.
Most work that we came across in computer architecture literature takes this approach.
On one hand, this approach leaves the system in an easily comprehensible and easily reproducible state.
On the other hand, performance results may be unrepresentatively better than they would be on a production system with fragmentation.

A second approach is to allow machines to become fragmented ``naturally'', most often by not restarting machines for period of time or between experiments~\cite{phamCoLTCoalescedLargeReach2012,gormanSupportingSuperpageAllocation2008,gormanWhatWhyWhere2006,yanTranslationRangerOperating2019}.
These approaches recognize the importance of fragmentation to system performance and may produce somewhat realistic fragmentation.
However, they are difficult to reproduce because the system state is highly dependent on what was run on the test machine.

Thirdly, and most commonly, prior work often runs artificial microbenchmarks or userspace applications that are specifically designed to induce fragmentation.
Most such approaches go through one or more cycles of allocating and deallocating anonymous memory~\cite{kwonCoordinatedEfficientHuge2016,michailidisMEGAOvercomingTraditional2019,bhattacharjeeTranslationTriggeredPrefetching2017,coxEfficientAddressTranslation2017,alvertiEnhancingExploitingContiguity2020,yanTranslationRangerOperating2019}.
Some approaches attempt to also introduce file-cache-based or kernel-pinned-memory-based fragmentation by including file-backed memory in these cycles~\cite{panwarHawkEyeEfficientFinegrained2019,panwarMakingHugePages2018,zhuComprehensiveAnalysisSuperpage2020}.
These approaches leave the system in a reproducible state before running experiments -- the cycle of allocations and deallocations can be reproduced by running a script.
However, like ``natural'' fragmentation approaches, it is unclear what that system state actually is or if it is representative of production systems; no prior work validates their methodology or justifies the number or type of allocation/deallocation cycles they use.

For this reason, we set out to create a new tool, \syssumf{}, with the following requirements:
\begin{itemize}
  \item Accurate: \syssumf{} should be able to artificially fragment physical memory such that end-to-end performance of workloads on the system is similar to if the system naturally reached a fragmented state.
  \item Science-friendly: \syssumf{} must produce easily interpretable and repeatable results.
  \item Low overhead: \syssumf{} should not require extraordinary amounts of hardware resources and should run quickly without adding overhead to workloads or system behavior.
  \item Easy to use: ordinary system developers and researchers should be able to use \syssumf{} without special expertise about fragmentation or the tool.
  \item Highly configurable: \syssumf{} should be flexible and able to adapt to new workloads, systems, or experimental environments.
  \item Does not require kernel modifications (but can be a kernel module): Direct kernel changes can be invasive, require maintaining a fork of the kernel, and are often difficult to port from version to version of the kernel. This requirement makes the tool significantly easier to use and maintain.
\end{itemize}

Our key idea in designing \syssumf{} is to summarize system fragmentation patterns using Markov Processes (MPs), which are a well studied and easily analyzed kind of \textit{stochastic process} -- a mathematical tool for describing probabilistic sequences.
\syssumf{} can use an arbitrary MP provided by a user to artificially (and repeatably) fragment a freshly booted machine.
MPs can compactly represent sequences that do not rely on extensive ``history'' -- that is, the transition from the current state to the next state depends only on the current state and its transition distribution (e.g., 50\% chance of jumping to state X, 20\% state Y, 30\% state Z).
Our intuition is to reproduce the effects of fragmentation (i.e., memory usage patterns as represented by MPs) without reproducing the process of fragmentation (i.e., a precise ordering of memory allocation and deallocation), which can be complicated and time consuming.
Because MPs are ``history-less'' \textit{per se}, and we want to ignore fragmentation history, MPs can efficiently represent fragmentation patterns.
Moreover, MPs are well-studied, have nice mathematical properties, and can be processed relatively efficiently.
Our hope was that \syssumf{} could be a foundation for improving experimental methodology and facilitating more scientific studies of memory management.

While our design achieved many of our goals, it ultimately failed at our first goal: accuracy.
In this section, we discuss the design of \syssumf{}, its merits and limitations.
While we do not recommend using \syssumf{} in its current form, we believe the ideas behind its design serve as a useful exploration of the design space and potential foundation for future work.

\section{Design and Implementation}

Given the importance of fragmentation to system behavior, prior work artificially fragments memory using a variety of unvalidated techniques, as previously discussed.
To improve upon these methodologies and improve future work in memory manager design, we designed and implemented \syssumf{}, a tool for artificially fragmenting physical memory according to user-specified parameters.
\syssumf{} is designed to be fast, easy-to-use, and configurable.

\syssumf{} pioneers a fundamentally different approach to artificial fragmentation.
Prior approaches attempt to coax the system into a fragmented state, such as by allocating and freeing memory, mapping and unmapping files, or forbearing to restart.
That is, they attempt to reproduce events that lead to fragmentation as a side-effect.
In contrast, \syssumf{} attempts to directly reproduce the end results of fragmentation: interleaved placement and usage patterns of physical memory.
Concretely, \syssumf{} directly manipulates page placement and usage through a kernel module according to a pattern specified by an MP.
This leaves the system in a well known, easily quantified/verified, and reproducible state.

\syssumf{} attempts to solve two basic problems: (1) succinctly capturing memory fragmentation patterns, and (2) reproducing them accurately and quickly on a fresh machine without attempting to reproduce the events that led to the fragmentation.
We observe that physical memory fragmentation is, by definition, characterized by the interleaving of differently used memory regions.
Thus, \syssumf{} solves problem (1) by encoding memory usage interleavings as Markov Processes (MPs).
\syssumf{} attempts (and fails) to solve (2) via a kernel module that takes all free memory from the kernel and frees that memory strategically as the user's test workload (and the kernel itself) request memory.
Notably, \syssumf{} attempts to do so without reproducing the temporal ordering of memory allocations, so as to avoid slow and tedious capture-and-replay infrastructure.

While \syssumf{} should not be used in its current form, our exploration of the design space is instructive and can be the basis for future work.
A working artificial fragmentation tool can aid the evaluation of future system memory management software by giving researchers a validated way to artificially fragment memory, creating a more realistic environment in which to run system evaluations.
Our tool is built for Linux, but our ideas are easily extensible to other kernels, especially other open-source kernels which have similar memory management mechanisms, such as BSD.

\subsection{Usage and Target Setting}

\syssumf{} targets an environmental setting in which a new workload/system runs on an already fragmented machine in steady state.
This is a common environment in datacenters and high-performance computing, and recent work seemingly tries to replicate for evaluation purposes.

When using \syssumf{}, a workload is expected to run in two phases: (1) warmup and (2) observation.
In (1), the system is freshly rebooted and \syssumf{} is run with the given profile.
Then, the test workload is warmed up, causing the system to reclaim (now fragmented) memory from \syssumf{}.
In (2), the workload/system under test is run with instrumentation -- this is the phase for which results are reported.
For batch workloads, we run the workload once for warmup and again for observation.

\paragraph{Scope.}
\syssumf{} is inappropriate for some settings.

\textit{If the user's goal is to measure memory allocation costs during the warmup phase, then \syssumf{} is not appropriate.}
During the warmup phase, \syssumf{} acts in place of the kernel physical memory allocator by releasing fragmented memory back to the kernel under memory pressure, which it then uses to satisfy requests.
As a result, memory allocation costs (through \syssumf{}) are slightly more expensive than the kernel's allocator but significantly cheaper than compaction or swapping.

\textit{If the user's goal is to measure swapping or overcommitment costs, then \syssumf{} will interfere with results and is not an appropriate fragmentation methodology.}
We assume that the test machine is not heavily overcommitted and that the workload will mostly receive idle memory (rather than requiring reclamation).

\textit{If the user's goal is to measure the unpredictability of default NUMA behavior, then \syssumf{} is not appropriate.}
\syssumf{} always returns memory on the NUMA node requested by the allocator.
Because \syssumf{} captures all free memory on the system, it is able to repeatably return memory on the same node, unlike the standard kernel allocator.
Therefore, users of \syssumf{} should be aware of their workload's/system's NUMA preferences; to achieve correct and repeatable results, users should pin workloads to the same NUMA node and set the memory placement policy.

\subsection{Capturing Fragmentation Patterns}

\syssumf{} characterizes physical memory fragmentation by specifying the interleaving order of contiguous homogeneous memory regions spatially.
\syssumf{} focuses on capturing spatial, rather than temporal, patterns to avoid needing capture-and-replay infrastructure that related work in file systems requires~\cite{conwayFilesystemAgingIt2019,conwayFilesystemAgingIt2019,joukovAccurateEfficientReplaying2005,agrawalGeneratingRealisticImpressions2009,conwayFileSystemsFated2017,weissROOTReplayingMultithreaded2013,zhuTBBTScalableAccurate2005}, which are slow and require significant storage and processing for traces.

As in Section \ref{sec:fragstudy}, we rely on non-homogeneity as a proxy for fragmentation because homogeneity bounds contiguity.
Each region has a length $s$ in pages (up to 1024; larger regions are represented a sequence of smaller ones) and a usage $u$, which can be one of: file cache, anonymous, anonymous huge page, pinned/opaque, or free (these are the memory usages with the most impact on fragmentation according to Section \ref{sec:fragstudy}).
In other words, the region is characterized by a tuple $(s, u)$, which we call a memory \textit{class} for brevity.
Given a snapshot of an existing system (via
\codesm{/proc/kpageflags}), \syssumf{} breaks memory usage into a sequence of contiguous homogeneous regions, identifies each region's class, and characterizes the relative ordering of classes via a MP.

\paragraph{Background: Markov Processes.}
This section gives relevant (unrigorous) background information about Markov Processes.
Familiar readers may want to skip this section.

A Markov Process is a type of \textit{stochastic process}, a mathematical object representing a system whose state transitions follow a probability distribution.
Specifically, MPs are stochastic processes that obey the Markov Property: (informally) the probability of any state transition depends only on the current state and the (potential) next state, not any past steps the MP may have taken.

Intuitively, a MP can be though of as a directed (possibly cyclic) graph, where vertices represent states in the state space, edges represent possible state transitions, and edge labels represent the probability of a given transition.
That is, given two connected states $A$ and $B$, an edge $A \rightarrow B$ would be labeled by the probability that the MP transitions to $B$ given that it is in state $A$.
(The outgoing edges from any state $A$ must all sum to exactly 1).

An MP with a (finite) state space $S$ can be characterized fully by a unique $|S|\times|S|$ transition probability matrix $P$, where each element $P_{A,B}$ is the probabiliy of the transition $A \rightarrow B$.
This matrix has the convenient property that $P^n$ (i.e., $P$ raised to the power $n$) represents the probability of making any transition in $n$ steps.

In many cases, we can compute an MP's \textit{stationary distribution}, a sort of equilibrium point where the probability of being in any given state does not change from one time step to the next.
Intuitively, the stationary distribution represents the probability of being in any given state in the long run.

Computing the stationary distribution can be tedious/convoluted, but for many MPs we can take a shortcut.
We can compute the limit $\lim_{n \rightarrow \infty}P^n$, which expresses the probability of ending up in any given state starting from any given state in the long term.
If every state is reachable from every other state with non-zero probability (and the state space is finite and discrete), then this limit converges to the stationary distribution.
Intuitively, this is because, by the Markov Property, prior steps are forgotten, so if every state is reachable from every other state, we eventually forget where we started by the Law of Large Numbers.
(If every state is not reachable, we can get stuck in one part of the MP without the possibility of ever return to other parts of the MP, which complicates the math.
Such unreachable states are called \textit{transient}; more precisely, a state is transient if the probability of being in that state after infinite steps is 0).

We use these properties in our implementation of \syssumf{}.

\begin{figure*}[t]
\centering
\includegraphics[width=0.65\textwidth]{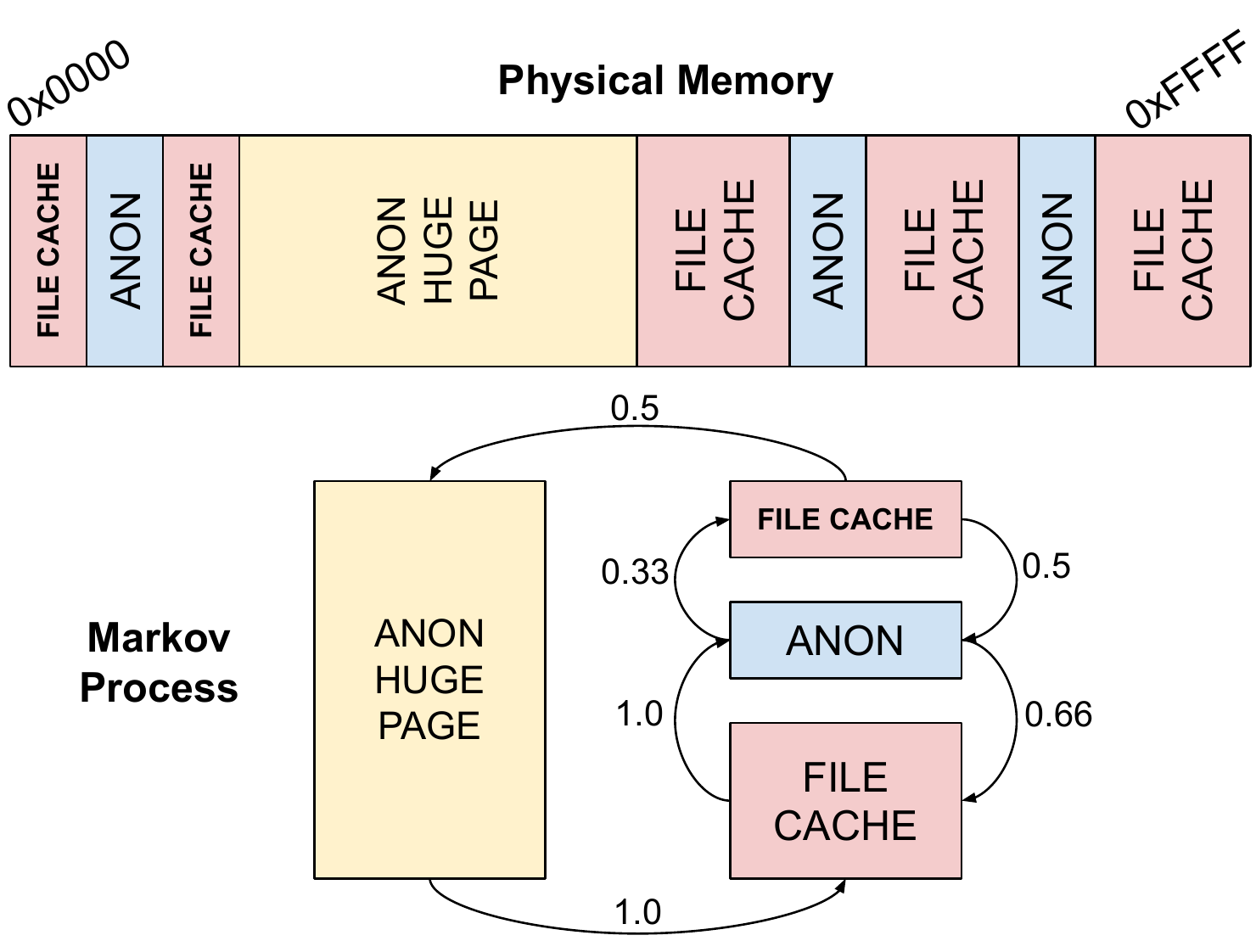}
\caption[Example: converting physical memory usage patterns to a MP]
{Example: converting physical memory usage patterns to a MP. Different sized-boxes represent different sized memory regions. \label{fig:exmp}}
\end{figure*}

\begin{lstlisting}[language=python, caption=Building an MP from \codesm{/proc/kpageflags} (pseudocode)., label=fig:mpbuild]
# Input: a sequence of homogeneous regions.
# Output: a fragmentation MP
def build_mp(regions):
  # P, the transition probability matrix.
  transitions = { }

  # For each pair of adjacent homogeneous regions: count
  # how often each $(size,usage)$ is followed by each 
  # other pair. Use to compute probabilities.
  for (s1, u1), (s2, u2) in sliding_window(regions, n=2):
    if (s1, u1) not in transitions:
      transitions[(s1, u1)] = { }
    if (s2, u2) not in transitions[(s1, u1)]:
      transitions[(s1, u1)][(s2, u2)] = 0

    transitions[(s1, u1)][(s2, u2)] += 1

  # Now turn counts into probabilities.
  # P[i,j] = Probability(i -> j)
  P = Matrix(len(transitions), len(transitions))
  for (s, u) in transitions:
    total = sum(transitions[(s, u)].values())
    for (s2, u2) in transitions[(s, u)]:
      P[(s,u), (s2,u2)] =
          transitions[(s,u)][(s2,u2)] / total

  # Cleanup: remove low-weight edges.
  for i,j in dimensions(P):
    if P[i,j] < THRESHOLD: # set by user
      P[i,j] = 0

  # Cleanup: Remove transient nodes.
  reachability = compute_reachability(P)
  for state in reachability:
    if state is NotReachable:
      P.remove(state)

  # Cleanup: Make sure the whole graph is connected by 
  # adding back low-weight edges in both directions 
  # between all connected components.
  ccs = connected_components(P)
  if len(ccs) > 1:
    for comp1, comp2 in all_pairs(ccs):
      s1 = pick_random_state(comp1)
      s2 = pick_random_state(comp2)
      P[s1,s2] = 0.1 # an arbitrary small probability
      P[s2,s1] = 0.1

  # Make sure probabilities still sum to 1
  renormalize(P)

  return P
\end{lstlisting}

\paragraph{MP Formulation of Fragmentation.}
We represent fragmentation using MPs where each state represents a different $(size, usage)$ tuple, and each transition $(s_1,u_1) \rightarrow (s_2, u_2)$ represents a region of size $s_2$ and usage $u_2$ following a region of size $s_1$ and usage $u_1$ spatially in the physical address space.
Figure \ref{fig:exmp} gives a toy example MP.

Building such an MP is straightforward given \codesm{/proc/kpageflags}, which allows us to build profiles derived from real machines.
Figure \ref{fig:mpbuild} shows our algorithm.
We first process \codesm{/proc/kpageflags} into a sequence of homogeneous regions, as done in Section \ref{sec:fragstudy}.
Then, given a sequence of regions, we can efficiently find all unique $(size, usage)$ pairs and compute the frequency of each pair being followed by another.

Finally, we run a simple ``cleanup'' procedure on each MP to make it more compact and easier to analyze.
The cleanup first trims extremely unlikely edges from the MP.
This reduces the size of the MP and reduces noise due to unlikely events.
Then, we ensure that there are not multiple connected components in the MP by adding equal-weight edges connecting unconnected components.
Finally, we trims any states that are transient.
These last two steps make it easy to compute the stationary distribution, which we use in our evalutation.

\paragraph{Rationale: Why MPs?}
When designing \syssumf{}, we needed a format to capture and concisely record interleaved memory usage patterns.
In accordance with our goals for \syssumf{}, the format must be robust in a number of ways.
Markov Processes satisfy these requirements.
In contrast, prior methodologies are based off of intuition that running some sequence of operations/processes will jumble up free pages, cause files to enter the file cache, or pin kernel data structures in fragmentation-inducing places, but they provide no guarantees about or quantification of the resulting system state.

First, MPs are efficient but expressive.
The size of the MP is bounded by the size of the state space.
We chose a state space of $\{free, file, anon, anonthp, opaque/pinned\} \times \{1,2,...1024\}$.
This is large enough to express many useful fragmentation patterns (over 26 million possible pairs of adjacent memory regions) but small enough to be compact and fast to compute over.
Our algorithms require polynomial time and space with respect to the size of the state space or are linear with respect to the size of \codesm{/proc/kpageflags}.

Second, unlike, e.g., many machine learning models, MPs have intuitive meaning to humans.
States in the MP correspond closely with memory allocations and usage.
Edges in the MP indicate interleavings between memory regions.
Studying these patterns may lead to insights about pathologies of MM implementations, such as if one type of memory usage is more prone to be broken up due to reclamation.
Moreover, human-readable formats are easier to debug, aiding our implementation of \syssumf{}.

Third, MPs have numerous convenient mathematical properties.
MPs are well defined and well studied, so we can specify fragmentation patterns precisely.
We can use well known methods to compute their properties, such as using the stationary distribution to find the expected amount of memory with different usages or the median contiguity of memory.
These properties make it possible to rigorously test and validate the (in)correctness of \syssumf{} -- we can check the outcome (e.g., from \codesm{/proc/kpageflags}) against the mathematical prediction.
Moreover, because MPs follow well defined rules, we can be confident that the results of \syssumf{} are repeatable.
Thus, MPs significantly enhance the science-friendliness of \syssumf{}.

\subsection{Reproducing Fragmentation Patterns}

\begin{lstlisting}[language=python, caption=Fragmenting a system based on a MP (pseudocode)., label=fig:mppseudo]
def fragment_node(node, npages, mp):
  # Capture all memory on the given NUMA node.
  captured_pages = alloc_pages_node(node, npages)

  # Lists of pages that have been assigned a usage.
  pages_file = []
  pages_anon = []
  pages_anon_huge_page = []
  pages_pinned = []

  # Fragment memory
  mp_node = first_node(mp)
  while not is_empty(captured_pages):
    (size, usage) = mp_node

    assigned_pages = remove_n_pages(captured_pages, size)

    switch usage:
      FILE: pages_file.append(assigned_pages)
      ANON: pages_anon.append(assigned_pages)
      ANONHP: pages_anon_huge_page.append(assigned_pages)
      PINNED: pages_pinned.append(assigned_pages)
      FREE: free(assigned_pages)

    mp_node = mp_rand_step(mp, mp_node)

  # Randomize lists to make shrinker efficient
  randomize(pages_file)
  randomize(pages_anon)
  randomize(pages_anon_huge_page)
  randomize(pages_pinned)

def mp_rand_step(mp, current):
  outgoing_edges = current.edges
  probabilities = current.edges_weights
  edge = select_random(outgoing_edges, probabilities)
  return mp.nodes[edge.to]
\end{lstlisting}

Using the \syssumf{} kernel module, a fresh system can be fragmented to reproduce the memory usage patterns from an MP.
At a high level, \syssumf{} captures all free memory on the system and releases strategically as the kernel and userspace request memory.
This allows \syssumf{} to control the placement of those allocations and produce the desired fragmentation patterns.

First, \syssumf{} takes all free memory away from the standard physical memory allocator.
On a freshly rebooted and idle system, this will be nearly all memory in the system.
Second, \syssumf{} does a random walk over the given MP: starting from a random node $A$, a random outgoing edge $A \rightarrow B$ is chosen according to the edge probabilities, and the class of $B$ becomes the next step (i.e., class) in the random walk.
For each class $(size, usage)$ -- each step in the random walk -- \syssumf{} sets aside $size$ pages for usage $usage$.
The random walk proceeds until all pages \syssumf{} took from the kernel have been set aside for some usage.
The pages set aside for each usage are kept in a linked list, but are not freed back to the system yet.
Note that this design naturally allows scaling fragmentation patterns to any size machine by changing the length of the random walk.

At this point, the system is ready to run the user's \textit{warmup} workload.
As the workload consumes more memory, \syssumf{} will relinquish memory back to the kernel's free memory pool from the appropriate per-usage linked lists, as discussed in the next section.
The pages released by \syssumf{} will be used by kernel and userspace allocation requests, which gives \syssumf{} the ability to control page placement.
By releasing memory from the appropriate linked lists (which were established by the MP), \syssumf{} ensures that the patterns established by the MP are preserved in the placement of kernel and userspace allocations, even after the memory leaves \syssumf{}'s direct control.
In contrast, prior methodologies often release all memory before the test workload run, giving free adjacent pages a chance to coalesce and defragment.

\paragraph{Shrinker.}
To gradually release memory while preserving fragmentation, \syssumf{} implements a \textit{shrinker} -- a mechanism in Linux that allows the swapping subsystem to request memory back from kernel modules (e.g., to implement a cache).
The shrinker can be invoked either synchronously (e.g., during a page fault) or asynchronously (e.g., via the swapping daemon).
\syssumf{}'s aforementioned per-usage linked lists are uniformly randomized after the random walk but before the user's workload runs; when a page needs to be released, the head of the list is chosen.
This simulates the effects of memory pressure and reclamation in Linux, which tend not to target pages in spatial order.

\syssumf{} releases pages marked as file and anonymous (non-huge-page) memory first, followed by anonymous huge page memory.
This follows our observation in Section \ref{sec:fragstudy} that file and anonymous (non-huge-page) memory become fragmented first and are targeted by reclamation first.
Simulated pinned/opaque pages are never released.

\paragraph{Limitation: Temporal Order.}
\syssumf{} does not attempt to reproduce the lifetime or temporal ordering of allocations.
Prior work in file system aging preserves the timing of allocation and reclamation, often with high fidelity, but at the expense of extremely complicated tool design and very high overhead/difficulty of use.
In contrast, \syssumf{} preserves the \textit{spatial} ordering of homogeneous regions and selects pages to be freed uniformly at random.
This may induce worse fragmentation than is realistic, as it will affect all parts of the physical address space, rather than particular high-churn areas of the address space.
We believe this is acceptable because it represents a worst-case scenario and tests the workload under observation under more extreme circumstances.

\paragraph{Limitation: Variation across Systems.}
\syssumf{} is primarily designed for and tested on Linux.
It is unclear whether other OS kernels experience similar levels and modes of fragmentation.
However, many other systems use similar memory management mechanisms, so we expect \syssumf{} can be extended for other systems easily.

\section{(In)Validating \syssumf{}}

When building \syssumf{}, a key goal was to produce a scientifically validated and reusable methodology: we wished to show that \syssumf{} realistically fragments physical memory according user specifications.
We use the data collected in Section \ref{sec:fragstudy} for our validation because it captures real system behavior.
A key challenge in validating \syssumf{} is the infeasibility of reproducing production workloads; we were unable to instrument the systems we observed and the workloads themselves are non-standard and non-repeatable anyway.

For this reason, we devised a two-step validation procedure for \syssumf{}:
\begin{enumerate}
\item Show that \syssumf{} can accurately reproduce fragmentation patterns recorded in \codesm{/proc/kpageflags}.
  In Section \ref{sec:fragstudy}, we collect \codesm{/proc/kpageflags} for all machines, giving us a real-world baseline.
  This step validates that \syssumf{} can reproduce realistic memory usage patterns.
\item Show that reproducing the fragmentation patterns recorded from \codesm{/proc/kpageflags} reproduces system-wide fragmentation effects.
  We fragment systems using \codesm{memhog} and a long-running YCSB workload~\cite{ycsb} and derive MPs from these fragmented systems.
  We measure the latency, throughput, and percentage of huge page memory for workload on these systems.
  Then, we reproduce the MPs and performance characteristics of the prior methodologies using \syssumf{} to show that end-to-end performance effects are reproduced by reproducing the MP.
\end{enumerate}

Our validation of \syssumf{} failed in step (2) above; we were unable to reproduce the end-to-end performance of workloads.
In this section, we present our methodologies and rationale in hopes that they can be useful to future research.

\subsection{Generated MPs}

We generate a profile (a MP) from the last snapshot of each instrumented machine during the observation window in Section \ref{sec:fragstudy}.
We use the last snapshot because it is the snapshot with the longest uptime for all machines observed and is most likely to represent a steady-state.

In general, we found the MPs were very complex and diverse.
Nearly every metric on which we tried to compare MPs had orders-of-magnitude variation between disparate MPs, including the number of nodes, the number of edges, the number of neighbors of similar type, the overall probability of a memory usage type, and the number of edges culled during MP cleanup.

A handful of MPs represented mostly idle machines -- these MPs generally had two states representing free and allocated memory.
However, the overwhelming majority of MPs had hundreds of nodes: most had between 300 and 800 nodes, but some had as many as 1650 nodes.
This indicates the sheer complexity of memory usage behavior in modern systems.

\subsection{Validating Fragmentation Patterns}

We show that in most cases \syssumf{} can accurately reproduce fragmentation patterns recorded in \codesm{/proc/kpageflags} from production systems, as specified by a fragmentation MP.
We quantify how well \syssumf{} reproduces the memory usage patterns captured by each profile $\mathcal{P}$ as follows:

\begin{enumerate}
    \item Use \syssumf{} to fragment memory according to $\mathcal{P}$.
    \item Collect \codesm{/proc/kpageflags} on the system after using \syssumf{}, using similar methodology to Section \ref{sec:fragstudy}.
    \item Compute the distribution $D$ of classes on the system (i.e., what fraction of memory falls in each class).
    \item Compute $S(\mathcal{P})$, the stationary distribution of $\mathcal{P}$, which is the ideal distribution over classes.
    \item Compare $D$ against $S(\mathcal{P})$ by computing an ``accuracy score'' as $\frac{\|S(\mathcal{P}) - D\|}{\|S(\mathcal{P})\|}$, the normalized difference between the empirical distribution $D$ and the theoretical ideal distribution $S(\mathcal{P})$.
\end{enumerate}

A rough (though flawed) intuition of the score is that it is the proportion of the time that \syssumf{}'s results differ from the ideal result.
The intuition is rough because the score can be $>1$ if $D$ contains a significant number of classes not in $S(\mathcal{P})$ (discussed more below).

A score of 0 indicates no difference and is a perfect (but unlikely) result.
For each profile, we use \syssumf{} to fragment 90\% of system memory; we found that 5-10\% of memory is already in use on a freshly rebooted system (e.g., by the kernel and background processes), and we make no attempt to reclaim or fragment this memory.
Thus, practically, a score of $\le 0.1$ represents an ideal result.

\begin{figure*}[t]
    \centering
    \includegraphics[width=\textwidth]{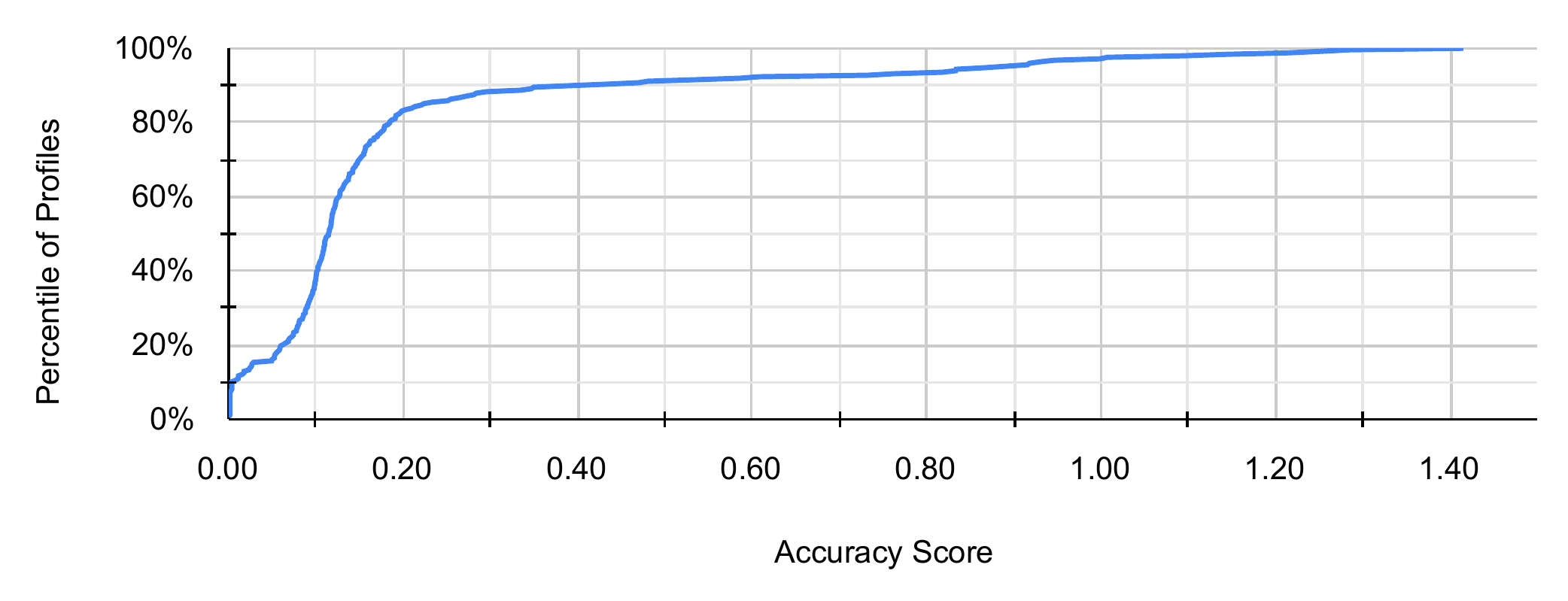}
    \caption{CDF of \syssumf{} accuracy scores for all machines. \label{fig:sumf-scores}}
\end{figure*}

\paragraph{Results.}
Figure \ref{fig:sumf-scores} shows the accuracy score for the patterns derived from the last snapshot of each of the 248 machines.
For 38\% of snapshots, \syssumf{} produces an ideal score of $\le 0.1$.
For 70\% of snapshots, \syssumf{} produces a score $\le 0.15$, within 5\% of ideal.
For 83\% of snapshots, \syssumf{} produces a score $\le 0.2$, within 10\% of ideal.
For 90\% of snapshots, \syssumf{} produces a score $\le 0.4$, within 30\% of ideal.

The remaining 25 out of 248 profiles score between 0.40 and 1.41, indicating poor reproduction of the profile.
We discuss these cases further in Section \ref{sec:sumfdisc}, along with implications about the use MPs for capturing fragmentation patterns.

\paragraph{Conclusion.}
In most cases, \syssumf{} is able to accurately reproduce fragmentation patterns derived from production systems.
Moreover, \syssumf{} users can easily measure their system to produce an accuracy score and check if their profiles are accurately reproducible before running experiments.

\subsection{Validating Performance \label{eval:complete}}

We measured the workload runtime, throughput, tail latency, peak memory usage, and amount of huge page usage for several benchmarks, using the same methodology as Mansi \textit{et al}~\cite{mansiCBMMFinancialAdvice2022}.
We used other artificial (not necessarily realistic) methodologies to fragment systems and measured their performance.
We then attempted to use \syssumf{} to reproduce the results we saw.
In particular, the two methodologies we used were (1) running \codesm{memhog}, and (2) running the \codesm{dc-mix} workload from our CBMM work~\cite{mansiCBMMFinancialAdvice2022}.

\begin{figure*}[t]
\centering
    \includegraphics[width=\textwidth]{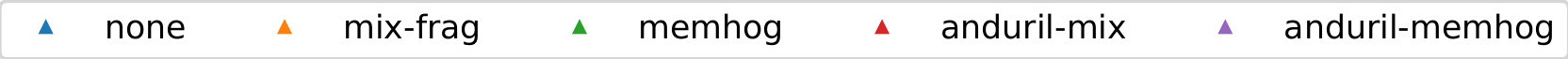} \\
    \subfloat[Runtime]{\includegraphics[width=0.24\textwidth]{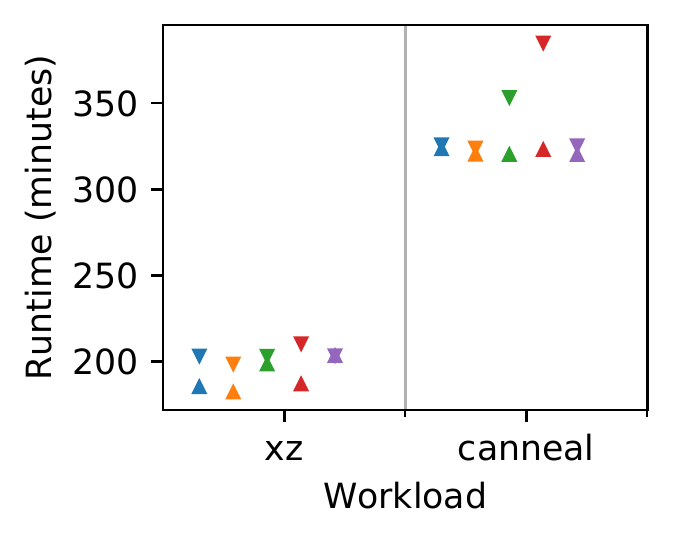}\label{fig:eval-rt}}
    \subfloat[Throughput]{\includegraphics[width=0.24\textwidth]{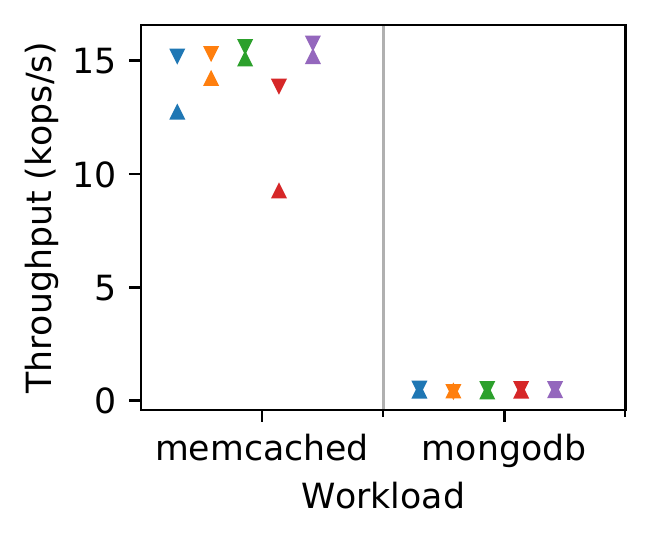}\label{fig:eval-tput}}
    \subfloat[Read p99 Latency]{\includegraphics[width=0.24\textwidth]{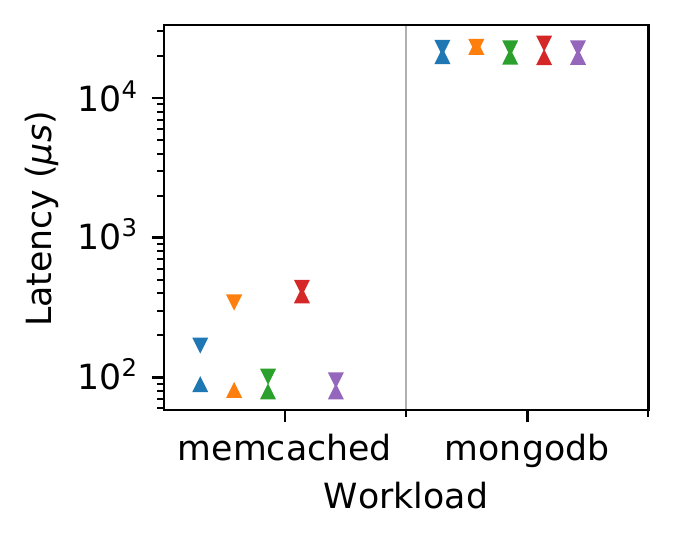}\label{fig:eval-rlat}}
    \subfloat[Update p99 Latency]{\includegraphics[width=0.24\textwidth]{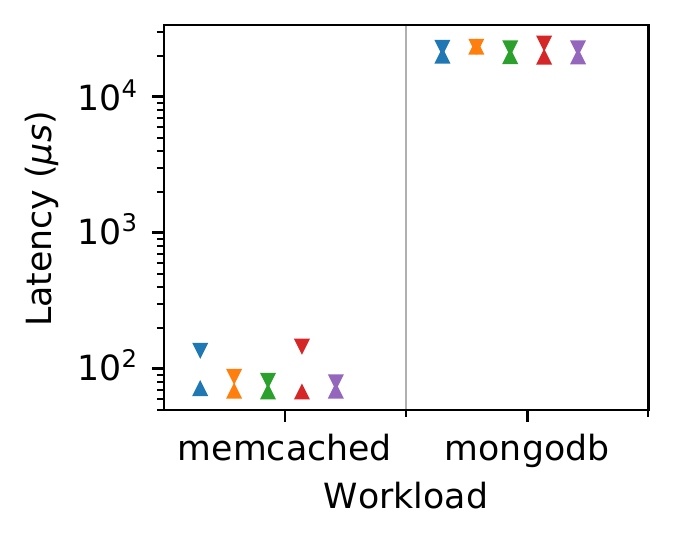}\label{fig:eval-ulat}} \\
    \subfloat[Peak Memory Usage]{\includegraphics[width=0.49\textwidth]{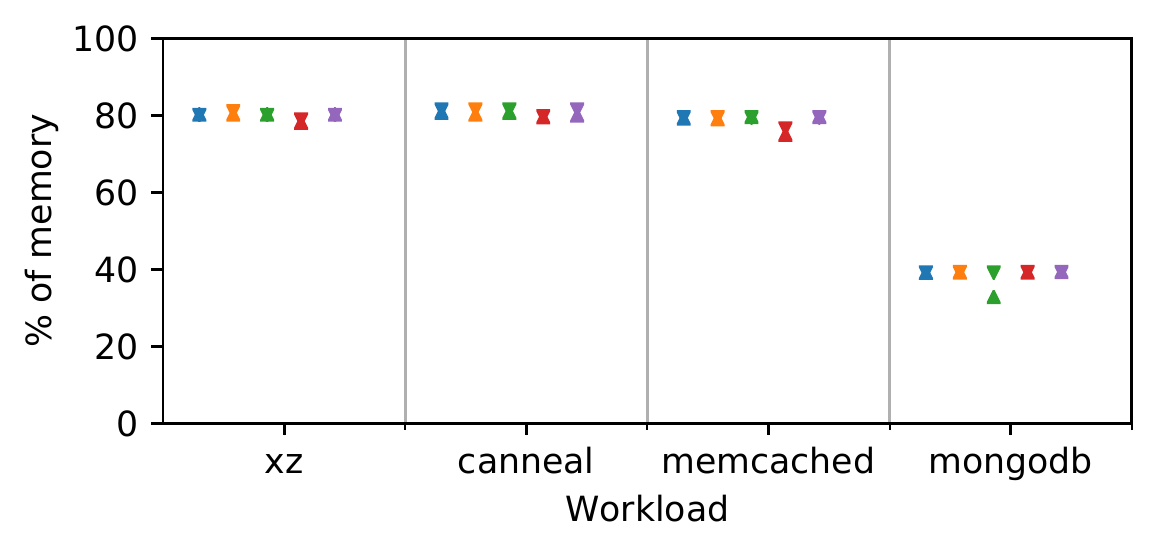}\label{fig:eval-peak-mem}}
    \subfloat[\% of Memory backed by Huge Pages]{\includegraphics[width=0.49\textwidth]{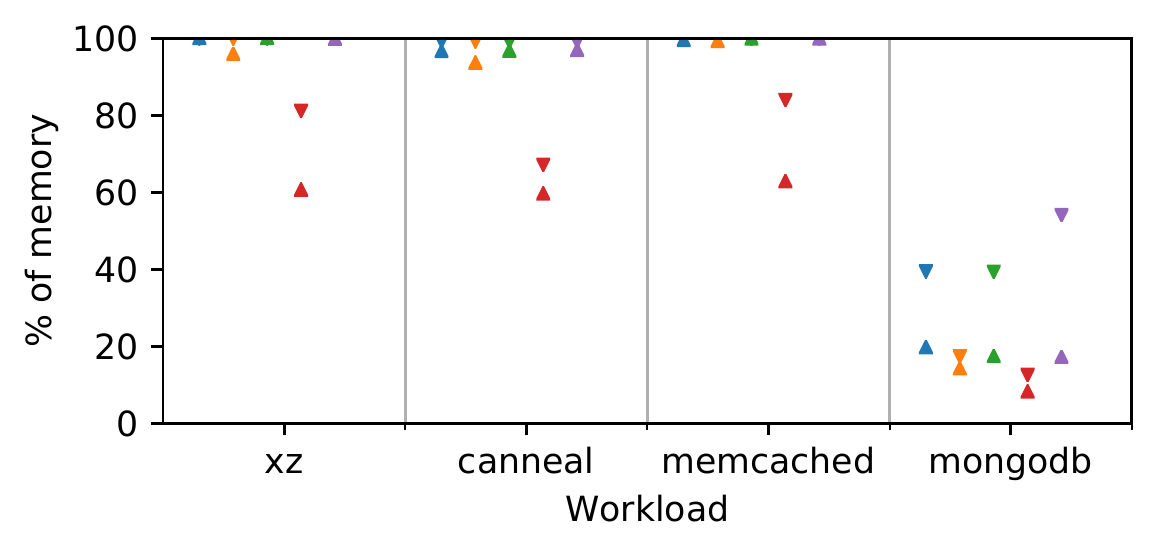}\label{fig:eval-hp}} \\
\caption[End-to-end performance with each fragmentation strategy]
{End-to-end performance with each fragmentation strategy. The upper and lower triangles indicate min and max values. ``none'' indicates no fragmentation (freshly rebooted system). \label{fig:sumfinvalid}}
\end{figure*}

\paragraph{Results.}
Figure \ref{fig:sumfinvalid} shows the results.
Ideally, ``\syssumf{}-mix'' (red) would match ``mix-frag'' (orange), and ``\syssumf{}-memhog'' (purple) would match ``memhog'' (green).

We found that for extremely simple MPs (e.g., all memory free), \syssumf{} is able to accurately reproduce end-to-end performance characteristics compared to a more ``naturally'' fragmented system; however, as previously discussed, simple MPs are rare in practice.
The ``\syssumf{}-memhog'' profile turns out to be such a profile, and in general the figure shows that the purple and green plots mostly match.
One exception is \codesm{canneal}, where \syssumf{} results in slightly different NUMA placement for data compared to Linux, leading to shorter worst-case runtimes in some cases.
We also found that sometimes \codesm{memhog} leads to slightly lower huge page usage for some runs of \codesm{mongodb} compared to \syssumf{}, leading to slightly lower peak memory usage (less memory bloat).

Unfortunately, in general, \syssumf{} produced end-to-end results inconsistent with the performance on the systems that generated the MPs used in our experiments.
In most cases, ``\syssumf{}-mix'' (red) differs ``mix-frag'' (orange).
``\syssumf{}-mix'' leads to significantly lower huge page usage in all workloads, leading to worse runtime/throughput/latency and slightly lower peak memory usage (less memory bloat).
Additionally, performance with ``\syssumf{}-mix'' was more variable than any of the other configurations -- the difference between max and min values was often an integer factor larger than ``mix-frag''.

\paragraph{Conclusion.}
\syssumf{} fails to preserve end-to-end system behavior for realistic MPs.
In the next section, we discuss potential reasons and the implications of \syssumf{}'s failure.

\section{Discussion: Accidental Failings and Incidental Findings\label{sec:sumfdisc}}

\syssumf{} fails to reproduce a minority MPs accurately.
Additionally, it fails to reproduce end-to-end system performance metrics.
In this section, we discuss these failures, attempt to learn from them, and present additional results with other \syssumf{} profiles.

\begin{figure*}
\centering
\includegraphics[width=0.4\textwidth]{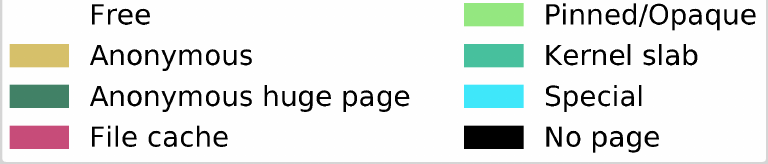} \\
\subfloat[Example Small VM]{\includegraphics[width=0.45\textwidth]{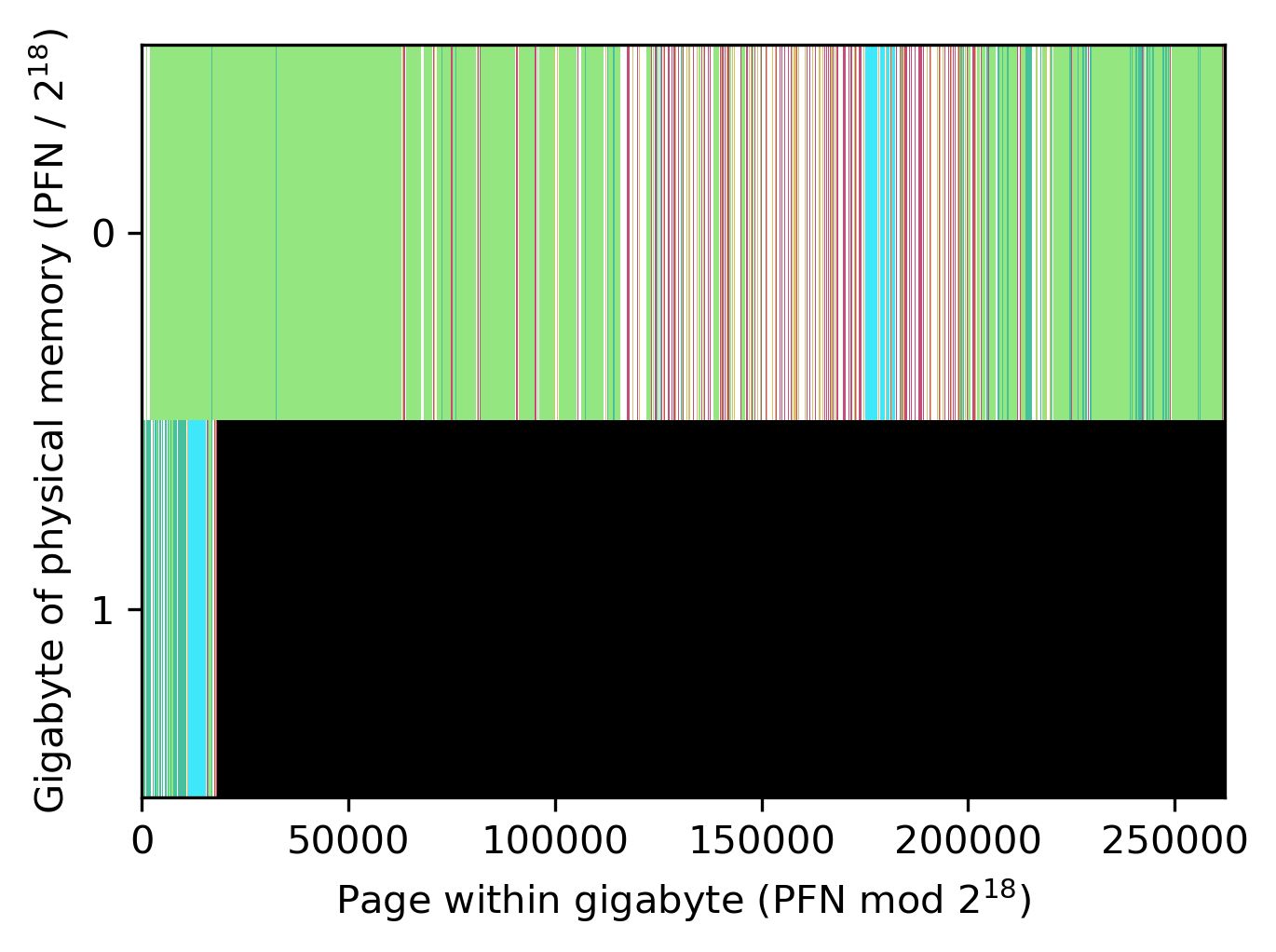}\label{fig:bad-cs220}}
\subfloat[Example Small VM with \syssumf{}]{\includegraphics[width=0.45\textwidth]{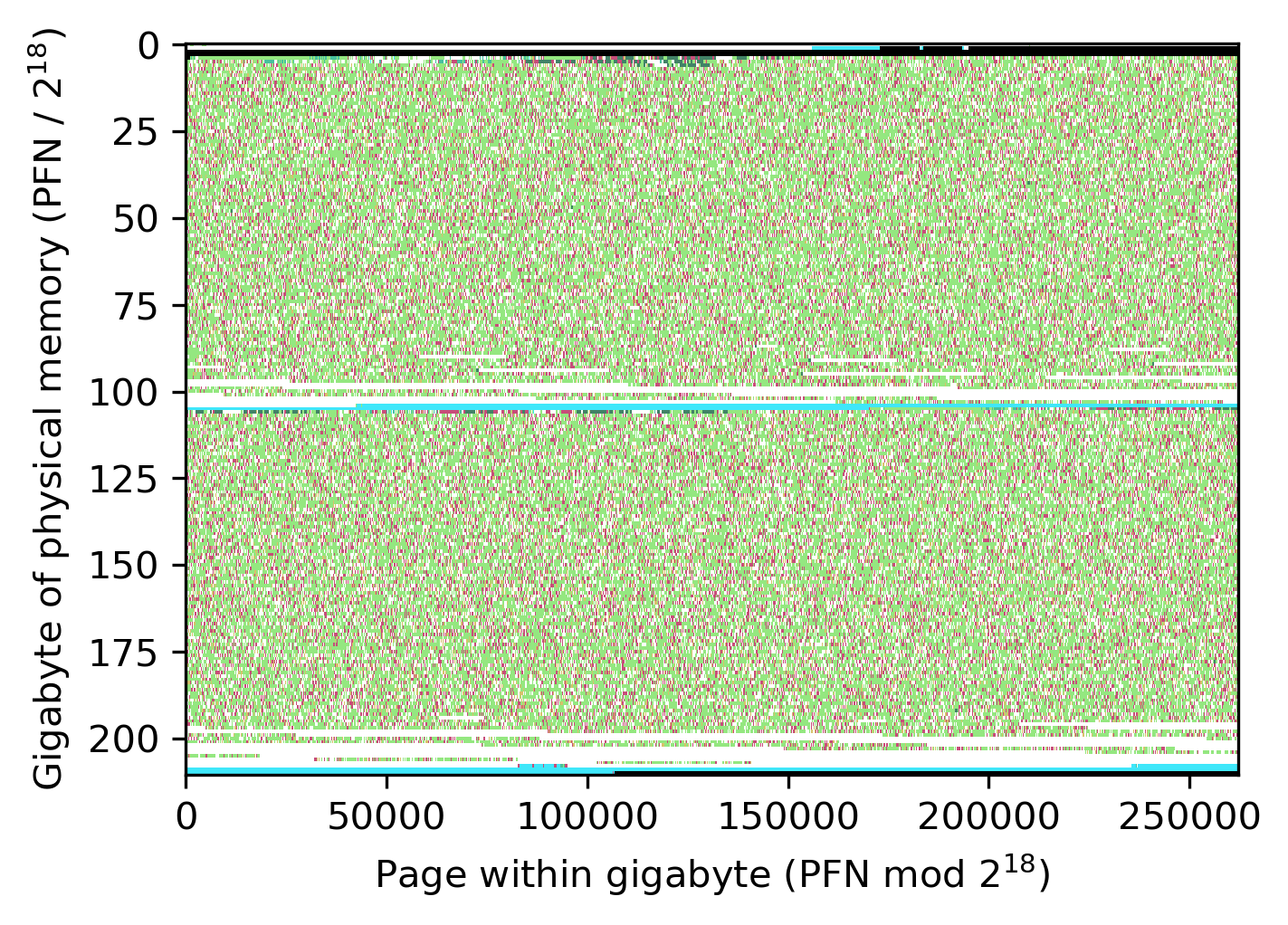}\label{fig:bad-cs220-andl}} \\
\subfloat[Example Monitoring Node]{\includegraphics[width=0.45\textwidth]{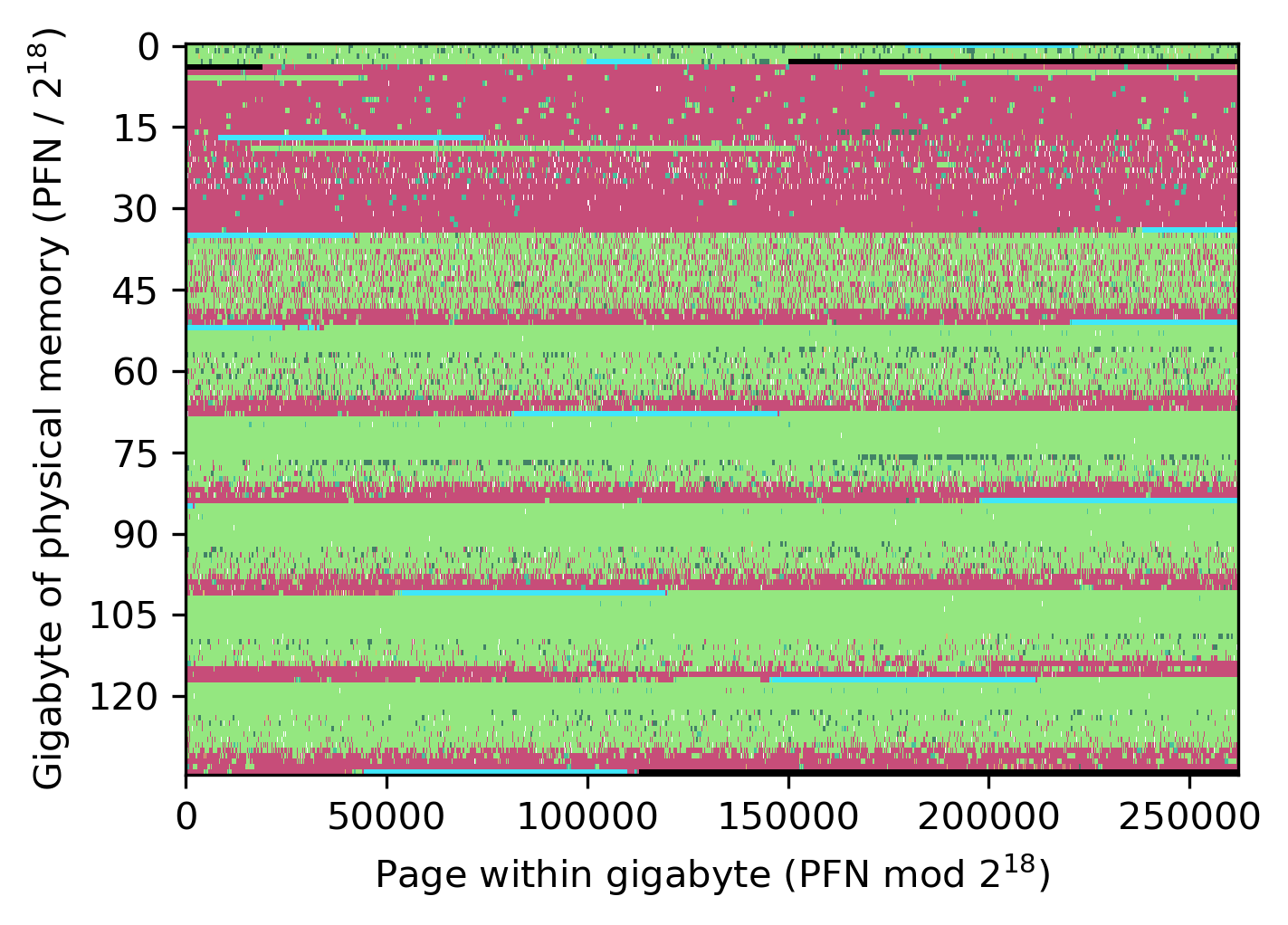}\label{fig:bad-graf}}
\subfloat[Example Monitoring Node with \syssumf{}]{\includegraphics[width=0.45\textwidth]{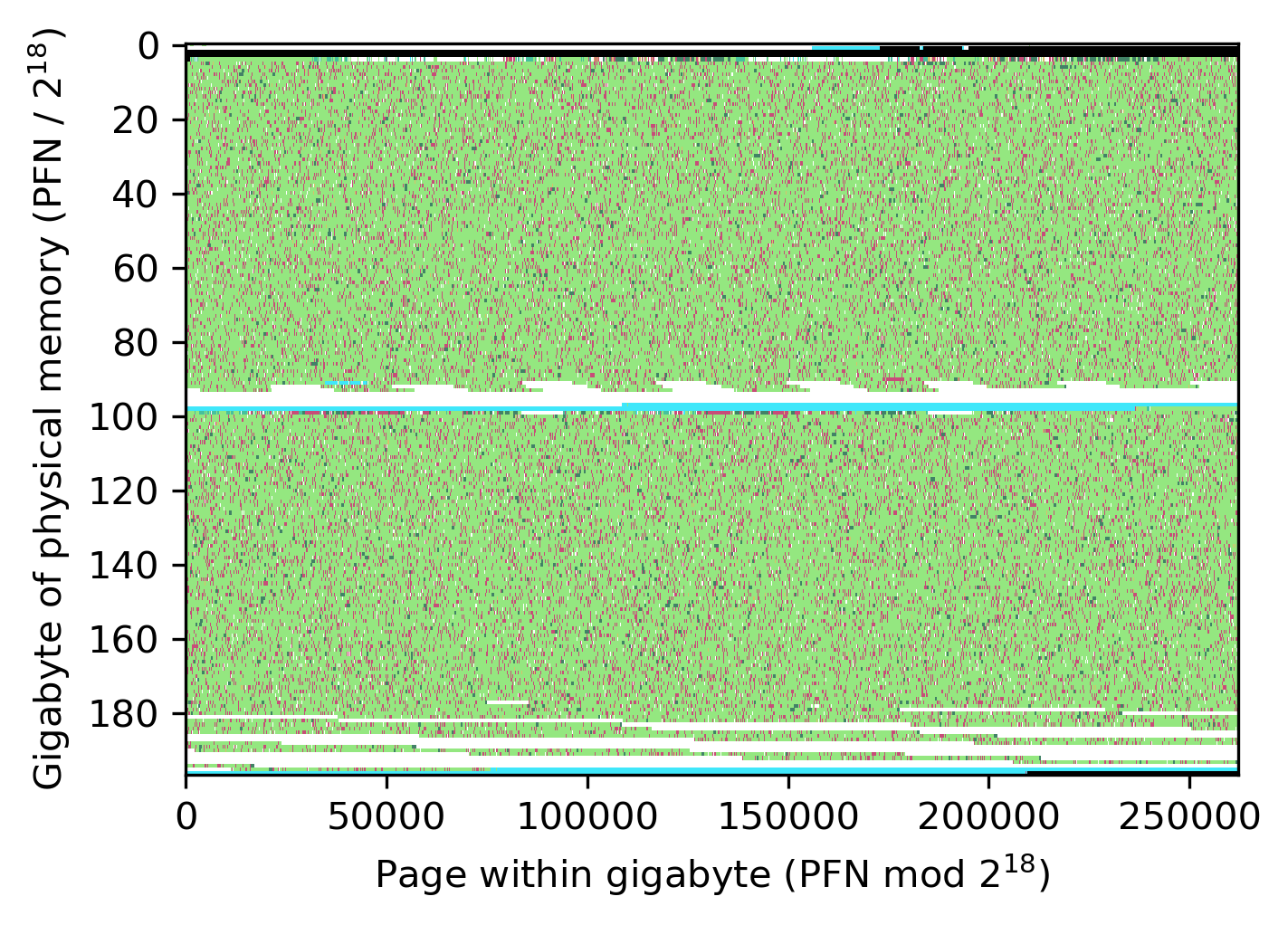}\label{fig:bad-graf-andl}} \\
\subfloat[Example Misc Node]{\includegraphics[width=0.45\textwidth]{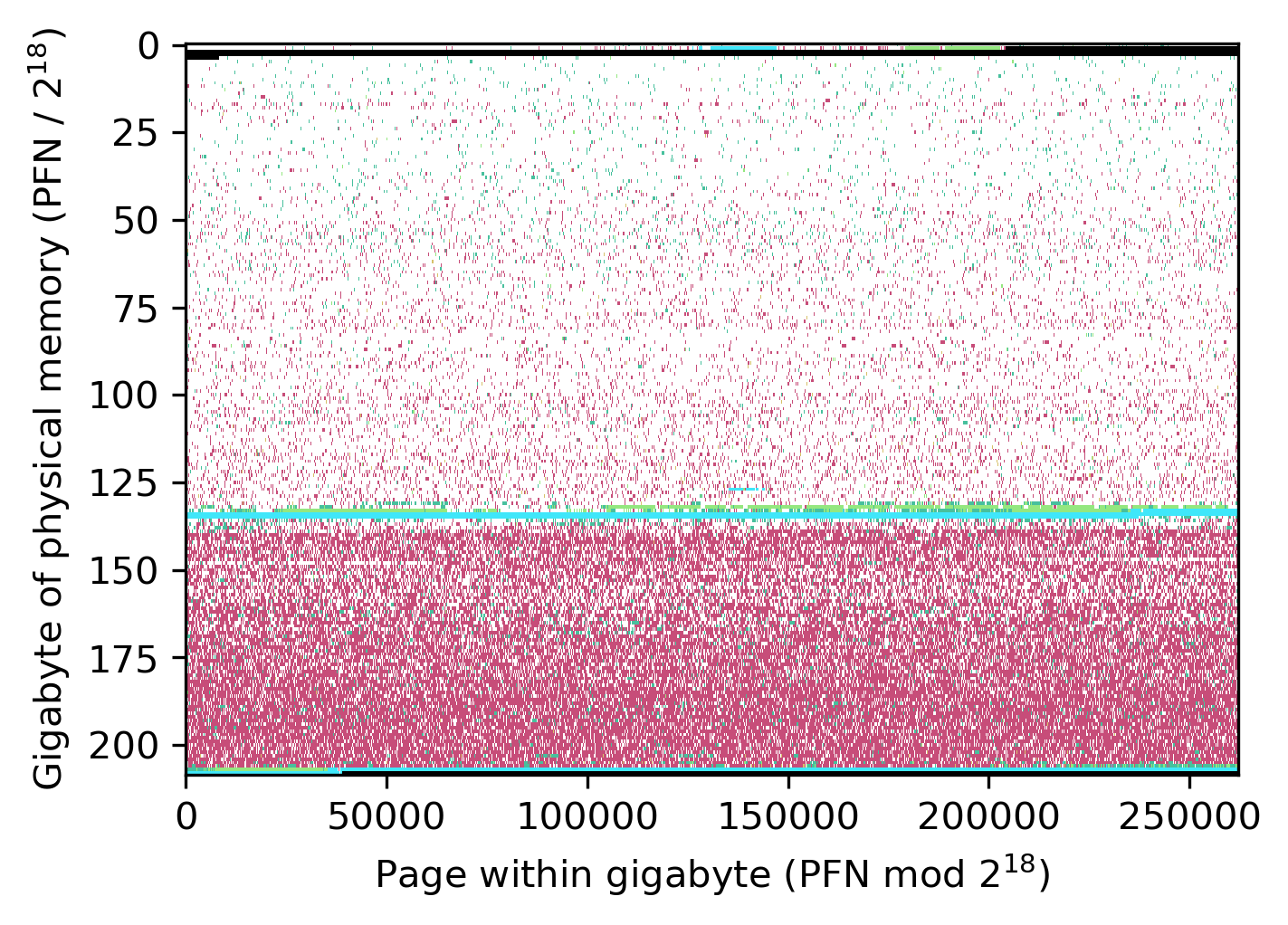}\label{fig:bad-bigben}}
\subfloat[Example Misc Node with \syssumf{}]{\includegraphics[width=0.45\textwidth]{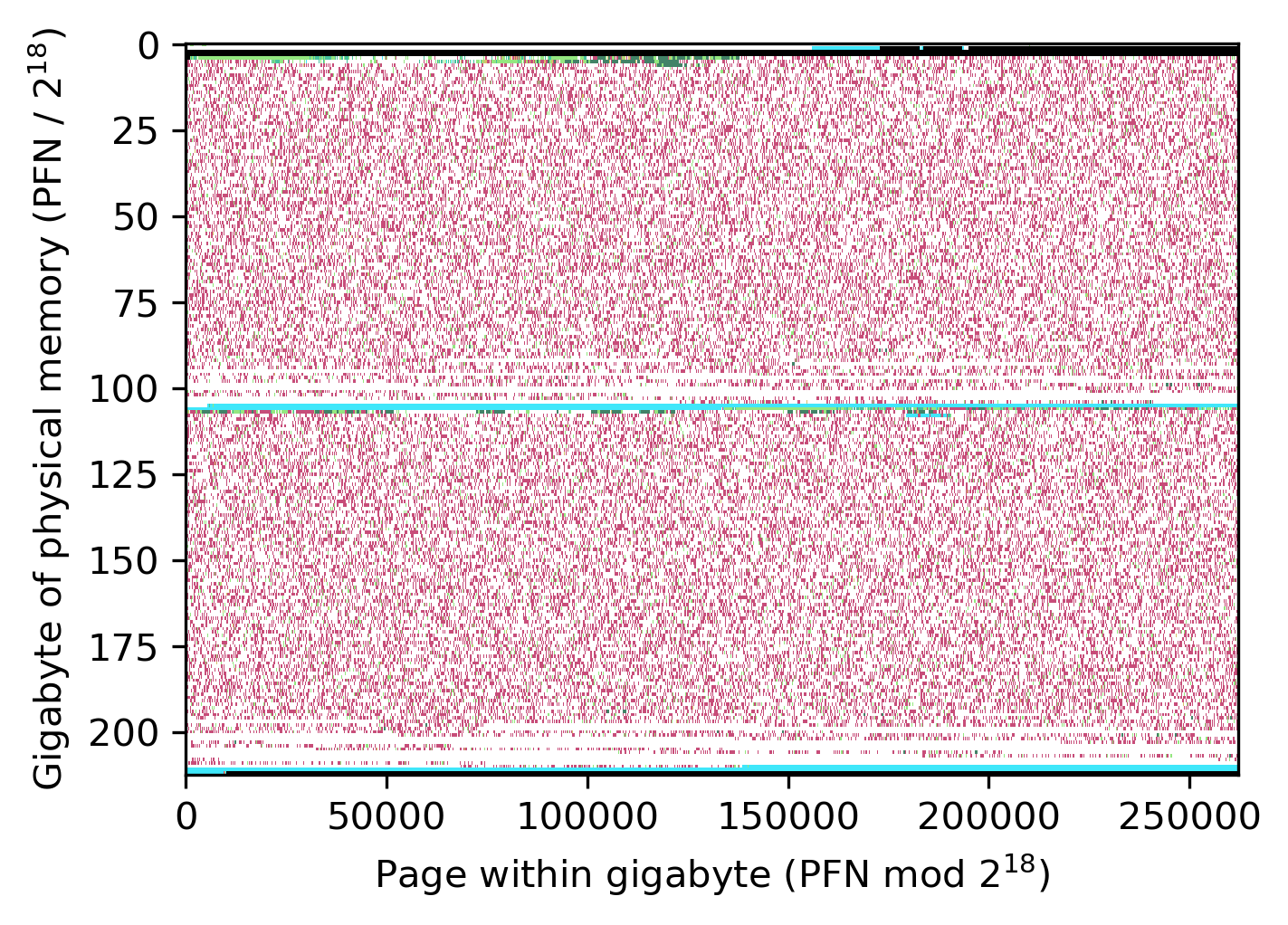}\label{fig:bad-bigben-andl}}
\caption[Example physical memory usage that generated MPs with bad accuracy scores]
{Example physical memory usage that generated MPs with bad accuracy scores. Each row represents a single gigabyte of physical memory, and each column represents a single 4K page within that gigabyte. Colors indicate what the page was used for. Plots on the left represent the memory usage on a real machine captured in Section \ref{sec:fragstudy}, while plots on the right represent \syssumf{}'s failed reproduction. \label{fig:bad-anduril}}
\end{figure*}

\begin{figure*}
\centering
\includegraphics[width=0.4\textwidth]{figures/bad-anduril-legend} \\
\subfloat[Example Hypervisor Node with Normal \syssumf{}]{\includegraphics[width=0.49\textwidth]{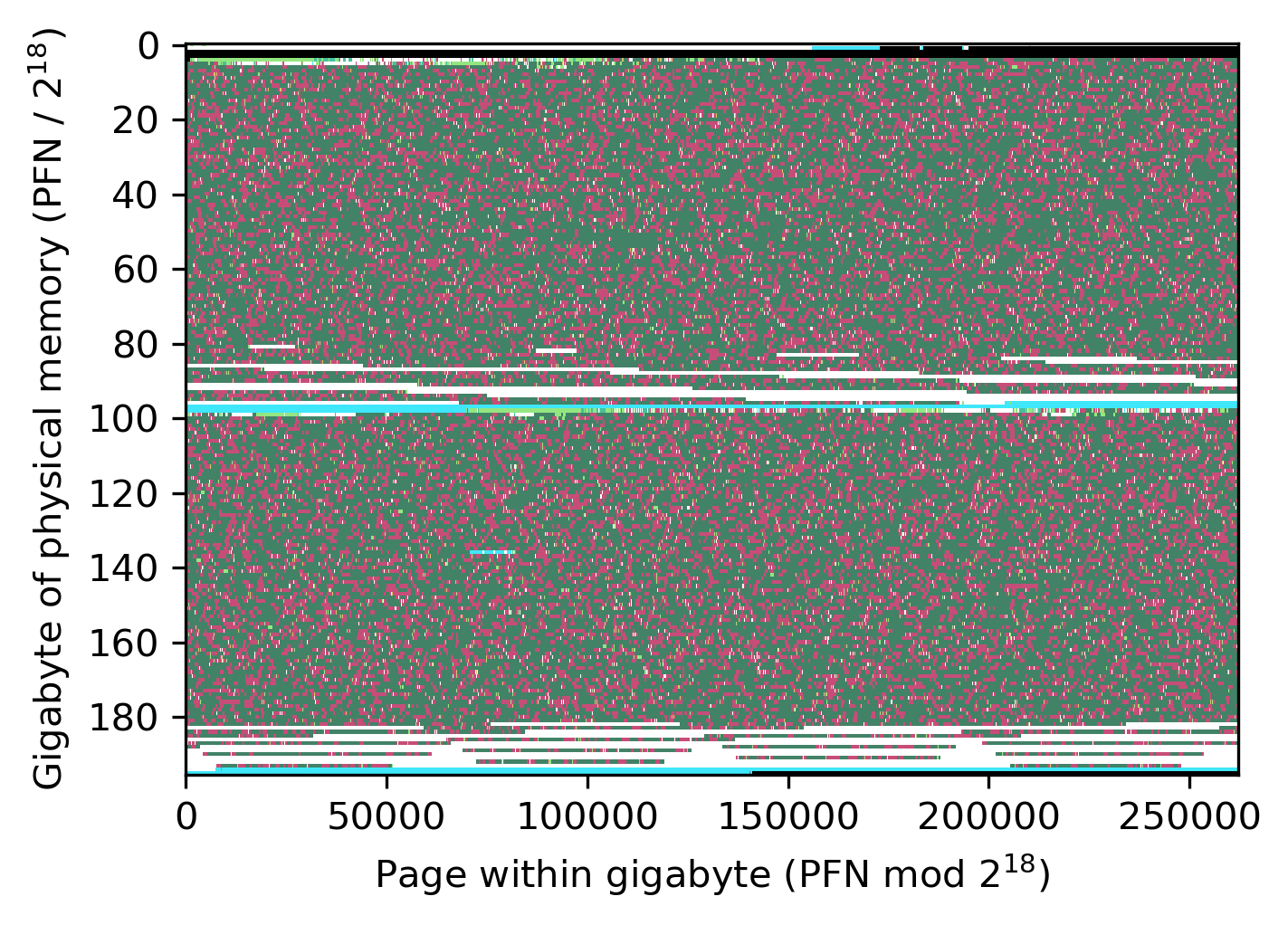}}
\subfloat[Example Hypervisor Node with Partition-Aware \syssumf{}]{\includegraphics[width=0.49\textwidth]{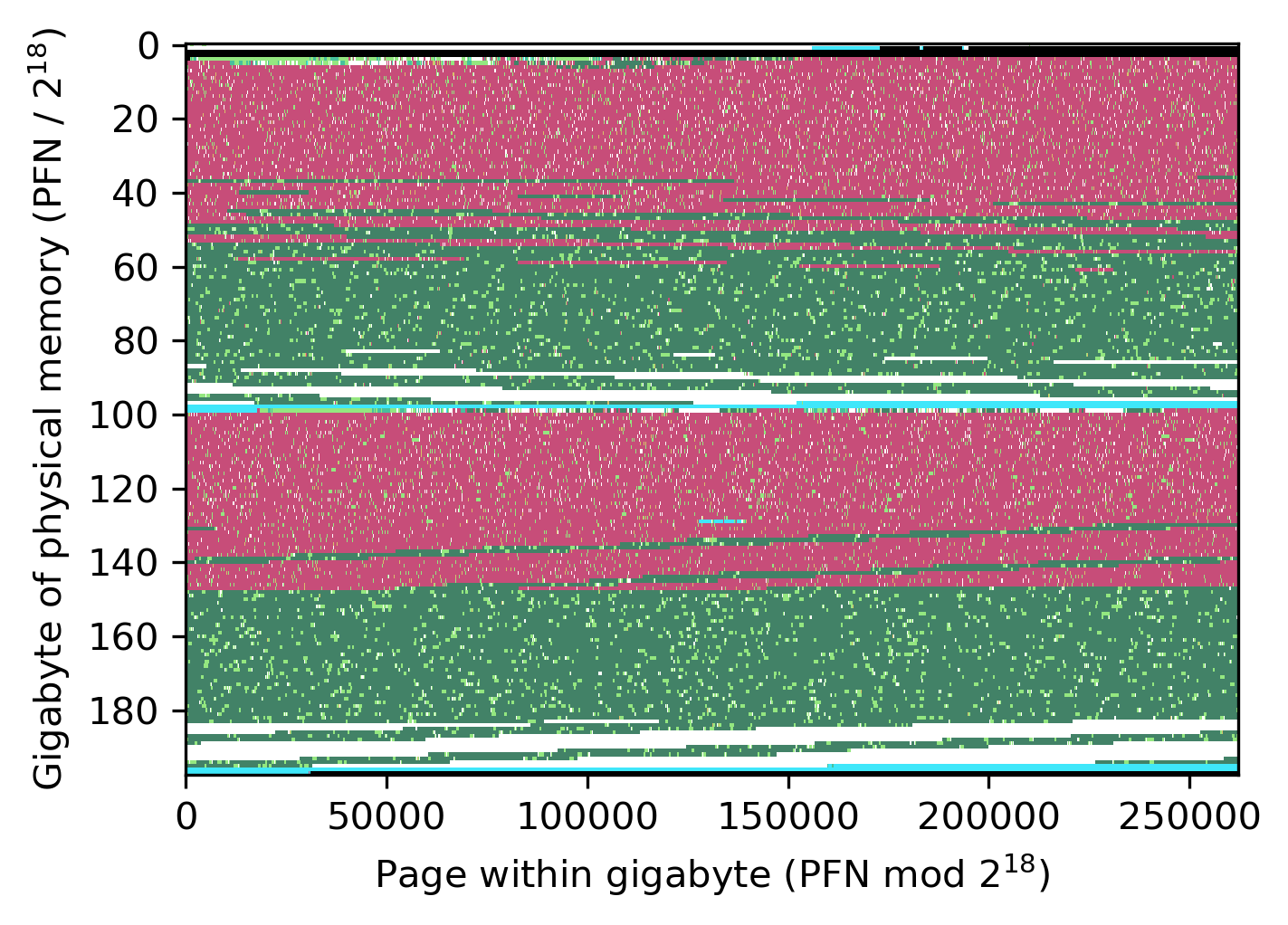}} \\
\caption[Example physical memory usage with Partition-Aware \syssumf{}]
{Example physical memory usage with Normal \syssumf{} and Partition-Aware \syssumf{}. Each row represents a single gigabyte of physical memory, and each column represents a single 4K page within that gigabyte. Colors indicate what the page was used for. \label{fig:hypermap-multi-part}}
\end{figure*}

\paragraph{Inaccurate Fragmentation Patterns.}
25 out of 248 profiles have an accuracy score $>0.4$ (with the highest being 1.41), indicating poor reproduction of the profile.
We made significant efforts to identify the causes of inaccuracy with mixed success.
We found that poor accuracy was statistically uncorrelated with the size (number of nodes or number of edges), shape (connectedness, number of similar neighbors, type of neighbors, or probability of types of memory usage), or starting point of the MP, all with $R^2 < 0.1$.
This could indicate a subtle bug somewhere in \syssumf{}, such as an incorrect assumption about the alignment or contiguity of memory regions; in our experience building  \syssumf{}, we found that such bugs are hard to find but have an outsized impact on accuracy.
However, we did identify one potential issue manifesting itself in two ways.

We call a usage pattern ``multi-partition'' if different regions of the address space exhibit radically different memory usage.
Because an MP does not track history, it does not easily express multi-partition patterns: the MP does not ``remember'' which partition it is in, so it tends to jump to the partition with the most frequent transitions overall, leading to a different overall mix of page usages than the machine originating the MP.
We observed this issue manifest in a few ways among the worst 25 profiles, as illustrated by Figure \ref{fig:bad-anduril}.

First, 18 of the 25 profiles came from machines that were one or more orders of magnitude smaller than our test machines.
In the extreme, several machines only had 1GB of memory, whereas our test machines had almost 200GB (we used the same 200GB machine for all experiments for the sake of consistency).
For many such machines, memory is strongly partitioned as the kernel claims one memory region and a small number of dominant applications claim the rest of memory.
Figure \ref{fig:bad-cs220} exemplifies one such machine, a small VM running a web server for a class; when \syssumf{} is run with its profile the result is Figure \ref{fig:bad-cs220-andl}.
Second, on machines with large amounts of memory, often memory usage changed for different regions of the physical address space.
We suspect, but cannot confirm, that the partitions correspond to different applications running on the machine.
Figures \ref{fig:bad-graf} and \ref{fig:bad-graf-andl} exemplify this pattern in an HTCondor health monitoring node.
In some cases, we observed that the different NUMA nodes had very different memory usage patterns, possibly as a result of NUMA placement policies.
Figures \ref{fig:bad-bigben} and \ref{fig:bad-bigben-andl} exemplify this pattern in a research node.
Notably, not all profiles with the above three patterns have inaccurate scores; many small machines and machines with clear partitions still had good accuracy scores.
We believe poor accuracy occurs when there is partitioning and memory usage patterns happen to result in an MP with very few transitions between one partition and the another.

Even when the accuracy score is good, it may be desirable to better represent multi-partition patterns.
There are natural extensions for MPs to allow them to track longer (but still finite) histories.
For example, the MP formulation can be changed so that each state contains two memory usages: the current one and the one from the step before, allowing us to look one step in the past.
This may be sufficient state to capture the needed patterns, but it also squares the number of MP states, making MPs significantly larger and more expensive computationally.
Another option is to add a layer of partitioning above the MPs: each profile would specify a number of partitions with a separate MP for each partition.
This approach is simple but has the disadvantage of complicating profile generation (how do we determine the boundaries between partitions?).
We did preliminary experiments with this approach by using two slightly modified copies of \syssumf{} running at the same time to reproduce a hypothetical 2-partition pattern with similar file and huge page usage as a hypervisor node.
Figure \ref{fig:hypermap-multi-part} shows an example of the results with a multi-partition pattern clearly visible, whereas with normal \syssumf{} a partition is not visible.

\paragraph{Inaccurate End-to-end Performance.}
As previously discussed, \syssumf{} does not accurately reproduce the end-to-end performance of different fragmentation profiles.
We suspect that part of the blame lies with our method of implementation: we found that Linux's \codesm{shrinker} mechanism seems to trigger after allocations fail at least once, leading the kernel to abandon normal allocation procedures.
In particular, as shown in Figure \ref{fig:eval-hp}, when \syssumf{} is used with a non-trivial profile (e.g., ``anduril-mix''), the amount of memory backed by huge pages is often drastically less, leading to worse runtime and tail-latency.
However, this does not explain all of the results we are seeing, as we have also tried experiments that do not rely on huge pages with similarly bad results.
We’ve also accounted for NUMA effects and designed our experiments to avoid measuring kernel allocator effects or noise.
One takeaway from this experience is that Linux does not currently have the requisite extensibility to control how pages are used without modifying the kernel itself.

Interestingly, our experiments to validate \syssumf{} not only invalidated \syssumf{} but also shined light on another potential concern: we found that running \codesm{memhog} before a workload does not significantly fragment a system.
After \codesm{memhog} terminates, its memory is freed and merged into large contiguous regions of memory.
In fact, if \codesm{memhog} causes memory pressure, the resulting reclamation and compaction can result in even greater contiguity after \codesm{memhog} terminates, leading to higher throughput and lower latency, as show in Figure \ref{fig:sumfinvalid}.
This finding highlights the need for carefully documented, reproducible, and validated fragmentation methodologies.

\begin{figure*}[t]
\centering
    \includegraphics[width=\textwidth]{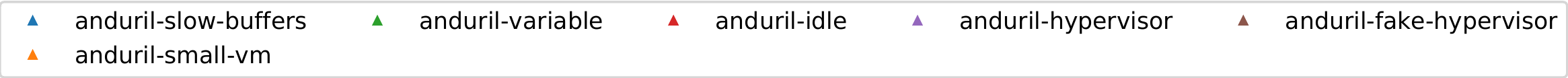} \\
    \subfloat[Runtime]{\includegraphics[width=0.24\textwidth]{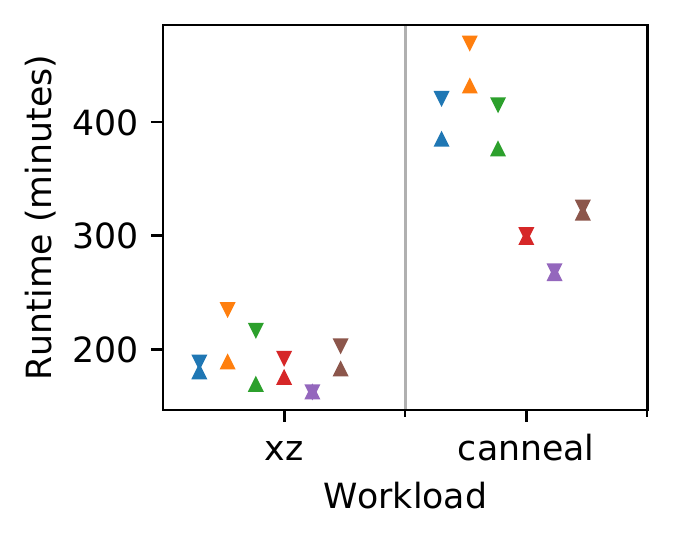}\label{fig:eval-rt-extra}}
    \subfloat[Throughput]{\includegraphics[width=0.24\textwidth]{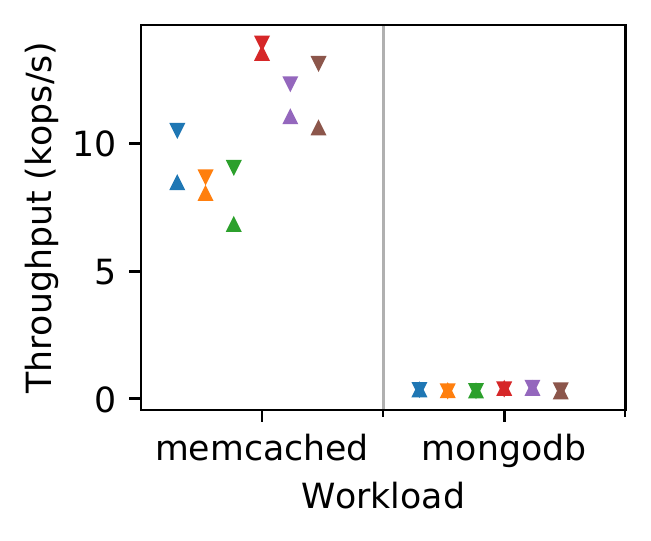}\label{fig:eval-tput-extra}}
    \subfloat[Read p99 Latency]{\includegraphics[width=0.24\textwidth]{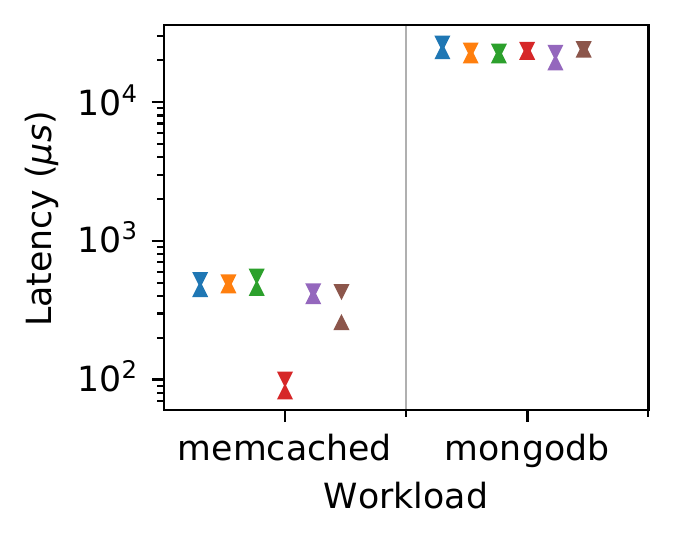}\label{fig:eval-rlat-extra}}
    \subfloat[Update p99 Latency]{\includegraphics[width=0.24\textwidth]{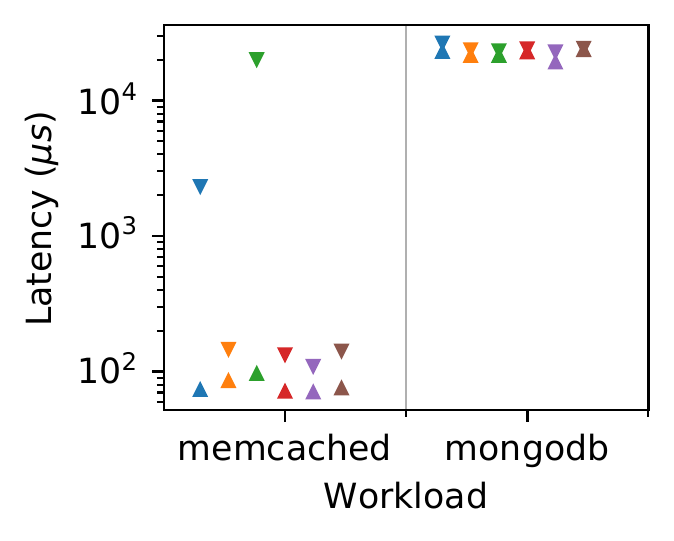}\label{fig:eval-ulat-extra}} \\
    \subfloat[Peak Memory Usage]{\includegraphics[width=0.49\textwidth]{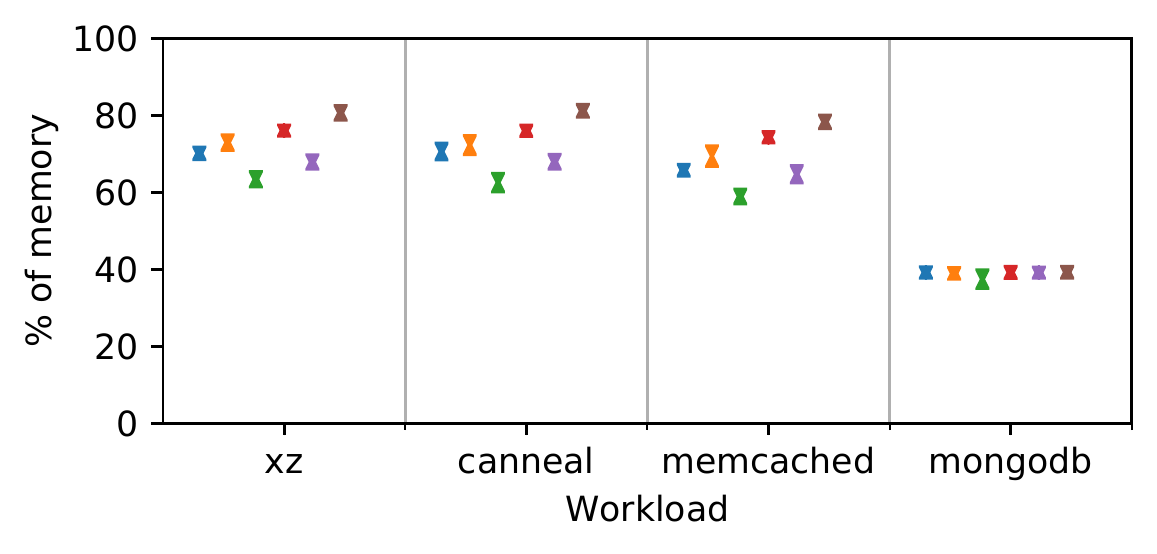}\label{fig:eval-peak-mem-extra}}
    \subfloat[\% of Memory backed by Huge Pages]{\includegraphics[width=0.49\textwidth]{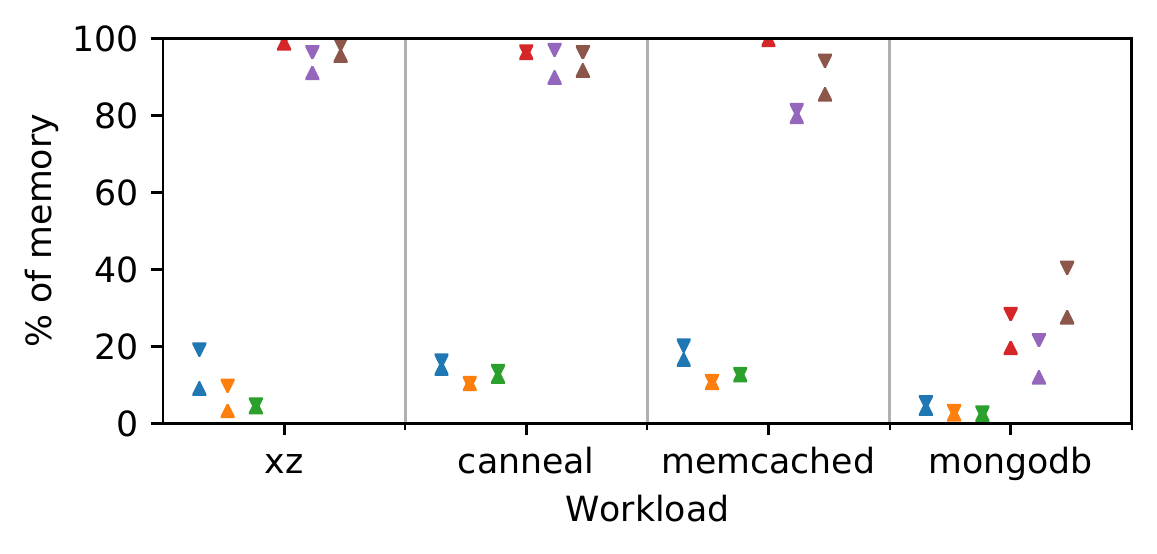}\label{fig:eval-hp-extra}} \\
\caption[End-to-end performance with different \syssumf{} profiles]
{End-to-end performance with different \syssumf{} profiles. The upper and lower triangles indicate min and max values. \label{fig:sumfinvalid-extra}}
\end{figure*}

\paragraph{\syssumf{}'s Effects.}
Even though \syssumf{} does not accurately reproduce the end-to-end performance of real systems, it may still shed light on interesting system behavior.
To this end, we run experiments with \syssumf{} using profiles derived from the memory usage patterns identified in Section \ref{sec:fragstudy} (Figure \ref{fig:exmemusage}), with two exceptions.
First, in the GPU-dominant memory pattern, almost all memory is dedicated to the GPU and is not usable for a memory-intensive workload; thus, it does not make sense to run our workloads with it, and we omit it from further results.
Second, we found that the hypervisor memory pattern suffers from the aforementioned multi-parition problem.
To achieve a balance of memory usage closer to what we observed on hypervisor machines, we manually created a new profile (called ``fake-hypervisor'').
We run experiments with both the hypervisor profile and fake-hypervisor profile.
For all profiles, we adjust the size of the workload so that its memory consumption fits in the memory leftover after kernel pinned pages and opaque (e.g., driver) memory usage is subtracted.
However, we run all workloads for the same number of operations where possible so that tail latency can be meaningfully compared.

Figure \ref{fig:sumfinvalid-extra} shows the results.
The ``slow-buffers'', ``variable'', and ``small-vm'' profiles were sharply separated from ``idle'', ``hypervisor'', and ``fake-hypervisor'' in almost all end-to-end performance metrics.
Figure \ref{fig:eval-hp-extra} reveals that a primary cause of the difference is the amount of huge page usage: the former three profiles used significantly fewer huge pages than the latter three, leading to longer runtime, lower throughput, and less memory bloat (lower peak memory usage).
Interestingly, most profiles had similar behavior with respect to tail latency, but latency was often significantly worse than in Figure \ref{fig:eval-hp}.
Possibly, \syssumf{} degraded tail latency enough that fragmentation effects were hidden.

We also note that ``fake-hypervisor'' had the highest huge page usage of any profile, which matches our observations of the hypervisor memory usage pattern in Section \ref{sec:fragstudy}.
Likewise, usage patterns with high degrees of fragmentation, such as ``slow-buffers'', ``variable'', and ``small-vm'', tended to have fewer huge pages and worse performance.
These results indicate that while \syssumf{} is not accurate, it does offer some control over fragmentation and its effects on system behavior, and those effects are reproducible and predictable.
This suggests that \syssumf{} could be useful to exercise different system behaviors even though it is inaccurate at reproducing realistic profiles.

\section{Conclusion}

More work is need to understand and mitigate fragmentation in current systems.
We observe several strong patterns that may be exploitable to get better performance or efficiency.
For example, while Linux supports a wide range of allocation sizes, only a small part of that range occurs commonly in practice.
Our work also suggests that huge page support in the file cache and fragmentation-aware reclamation algorithms could have a big impact to reduce fragmentation.

Additionally, while \syssumf{} does not accomplish the goals we intended in its current form, its design, implementation, and attempted validation are instructive.
Memory fragmentation behavior is surprisingly complex and hard to represent and reproduce efficiently.
MPs may be a useful means of doing so, but it is unclear if they capture all important information to reproduce the end-to-end performance effects of fragmentation.
Our findings suggest that \syssumf{} could be useful to exercise different system behaviors even though it is inaccurate at reproducing realistic profiles.
Significant further research is needed to develop and validate artificial fragmentation methodologies for future research.

Our attempt to build and validate \syssumf{} also reveals other interesting findings.
We found that Linux lacks key extensibility features in its MM subsystem that would make such testing and experimentation easy and efficient.
We had to resort to features designed for kernel modules with caches (i.e., \codesm{shrinker}), whereas what we wanted was extensibility of the memory allocator.
Meanwhile, current methodologies, such as running \codesm{memhog}, remain scientifically dubious -- depending on their exact usage, which is often insufficiently documented, they may have counterintuitive effects that make experimental results harder to interpret.

We hope that \syssumf{} can be a basis to make methodology more scientific, easier to use, and easier to understand.

\section*{Acknowledgments}

We thank CSL and HTCondor for allowing us to instrument their systems; in particular, many thanks go to Michael Gibson (CSL) and Aaron Moate (HTCondor), who put significant time and effort into helping us plan, build, and deploy the instrumentation; we could never have done this work without you.
We thank David Merrell for early insightful discussions about data analysis.
We also thank Jing Liu, Sujay Yadalam, and Anthony Rebello, for their insightful feedback on drafts of this paper, and the anonymous reviewers for their feedback on our submission.

This research was funded by NSF grants CNS-1815656 and CNS-1900758.

This research was performed using the compute resources and assistance of the UW-Madison Center For High Throughput Computing (CHTC) in the Department of Computer Sciences.
The CHTC is supported by UW-Madison, the Advanced Computing Initiative, the Wisconsin Alumni Research Foundation, the Wisconsin Institutes for Discovery, and the National Science Foundation, and is an active member of the OSG Consortium, which is supported by the National Science Foundation and the U.S. Department of Energy's Office of Science.

\clearpage
\bibliography{references}{}
\bibliographystyle{plain}

\clearpage
\appendix
\onecolumn

\section{Study Machines Details}

{\footnotesize
\begin{longtable}{|p{0.85in}|p{2.75in}|p{0.75in}|p{2.00in}|}
\hline
\bf Host Type (Amt)           & {\bf \footnotesize CPU Thds/Cores/Sockets (Family; N machines if >1)}
                            & {\bf \footnotesize Mem (N if >1)} & \bf Description \endhead \hline \hline
\texttt{hpc-exec}      (8 BM)  & 48T/48C/4S (Opteron'13), 32T/16C/2S (Sandy Bridge'12), 40T/20C/2S (Skylake'17), 40T/20C/2S (Ivy Bridge'13), 4T/4C/1S (HarperBridge'08)
                            & 128GB (5), 512GB (2), 16GB (1)
                            & Generic execution hosts \\ \hline
\texttt{hpc-gpu1}       (4 BM) & 32T/16C/2S (Haswell'14), 40T/20C/2S (Skylake'17), 96T/48C/2S (EPYC'19)
                            & 2TB (2), 512GB, 128GB
                            & GPU execution hosts \\ \hline
\texttt{hpc-gpu2}       (2 BM) & 160T/80C/4S (Skylake'17), 40T/20C/2S (Skylake'17)
                            & 512GB, 256GB
                            & Dockerized GPU workloads for particular project \\ \hline
\texttt{hpc-hmem}       (2 BM) & 80T/40C/4S (Ivy Bridge'14), 80T/40C/4S (Broadwell'16) &    2TB, 4TB & High-memory hosts \\ \hline
\hline  \pagebreak[3]
\texttt{hpc-hypr}       (6 BM) & 32T/16C/2S (Sandy Bridge'12), 12T/12C/2S (Opteron'10)
                            & 96GB (4), 64GB (1), 32GB (1)
                            & Hypervisor hosts \\ \hline
\texttt{hpc-ceph}       (2 BM) & 32T/16C/2S (Ivy Bridge'13)   &       128GB & Ceph object store hosts \\ \hline
\texttt{hpc-subm}       (2 BM) & 32T/16C/2S (Sandy Bridge'12) &        96GB & User access points for submitting jobs \\ \hline
\texttt{hpc-es}         (1 BM) & 12T/12C/2S (Opteron'10)      &        64GB & Host running elastic search + kibana \\ \hline
\texttt{hpc-graf}       (1 BM) & 48T/48C/4S (Opteron'13)      &       128GB & Host running graphana + nosql database \\ \hline
\hline  \pagebreak[3]
\texttt{hpc-fprx}       (2 VM) & 4 vCPUs (Intel, AMD)         &        32GB & Squid forward proxy \\ \hline
\texttt{hpc-rprx}       (2 VM) & 12 vCPUs (Intel)             &        32GB & Squid reverse proxy \\ \hline
\texttt{hpc-stag}       (1 VM) & 4 vCPUs (AMD)                &         8GB & Host for staging data into local HPC storage \\ \hline
\texttt{hpc-web}        (1 VM) & 4 vCPUs (Intel)              &         4GB & Static content web server \\ \hline
\texttt{hpc-dkca}       (1 VM) & 2 vCPUs (Intel)              &         4GB & Docker caching registry \\ \hline
\texttt{hpc-gang}       (1 VM) & 8 vCPUs (AMD)                &        64GB & Ganglia time-series metric gatherer \\ \hline
\texttt{hpc-mntr}       (1 VM) & 4 vCPUs (Intel)              &         8GB & Icigna/nagios monitoring service \\ \hline
\hline  \pagebreak[3]
\texttt{cs-rsch } (32~BM,~12~VM)     & 2 vCPUs (Broadwell '16; 7), 4T/4C/1S (Yorkfield '08; 6), 56T/28C/2S (Broadwell '16; 3), 2T/2C/1S (Blackford '05; 3), 1 vCPU (Broadwell '16; 3), 4T/4C/1S (Yorkfield '08; 2), 32T/32C/4S (Opteron '13; 2), 128T/64C/2S (EPYC '17), 128T/64C/2S (EPYC '20), 48T/24C/2S (Cascade Lake '19), 48T/24C/1S (EPYC '20), 40T/20C/2S (Skylake '17), 32T/32C/2S (Opteron '13), 32T/16C/2S (Skylake '17), 32T/16C/2S (Sandy Bridge '12), 32T/16C/2S (Haswell '14), 32T/16C/2S (EPYC '19), 24T/12C/1S (Ryzen '19), 16T/16C/2S (Opteron '13), 4T/4C/1S (Opteron '10), 4T/4C/1S (Ivy Bridge '12), 2C/2T/1S (Opteron '09), 15 vCPUs (Broadwell '16), 2 vCPUs (Haswell '14)
                  & 2.3TB (1), 1TB (1), 512GB (3), 384GB (1), 256GB (5), 225GB (1), 64GB (4), 33GB (1), 32GB (5), 16GB (1) 8GB (12), 4GB (3), 2GB (6) & Various hosts used by different research projects \\ \hline
\texttt{cs-misc } (23~VM,~6~BM)      & 1 vCPU (Broadwell '16; 8), 2 vCPUs (Broadwell '16; 7), 4 vCPUs (Broadwell '16; 5), 1 vCPU (Haswell '14; 3), 20T/10C/1S (Cascade Lake '19; 3), 64T/32C/2S (Cascade Lake '19), 2T/2C/1S (Blackford '05), 4T/4C/1S (Yorkfield '08)
                  & 192GB (1), 64GB (3), 32GB (1), 16GB (3), 8GB (5) 4GB (6), 2GB (10) & Misc hosts supporting specific use cases/projects \\ \hline
\texttt{cs-cuda }     (16 BM) & 48T/24C/1S (EPYC '19; 4), 48T/24C/2S (Broadwell '16; 3), 64T/32C/2S (EPYC '19; 2), 64T/32C/2S (Cascade Lake '19; 2), 48T/24C/2S (Cascade Lake '20), 96T/48C/2S (Cascade Lake '20), 48T/24C/2S (Cascade Lake '19), 36T/18C/1S (Skylake '17), 56T/28C/2S (Broadwell '16)
                  & 512GB (4), 480GB (1), 256GB (5), 192GB (2), 128GB (3), 96GB (1) & Research hosts using Nvidia CUDA~\cite{nickollsScalableParallelProgramming2008} \\ \hline
\texttt{cs-slrm }  (15 BM) & 48T/24C/2S (Haswell '14; 13), 32T/16C/2S (Skylake '17), 24T/12C/1S (Haswell '14)
                  & 384GB (1), 128GB (14) & SLURM compute nodes run on behalf of Stats dept \\ \hline
\texttt{cs-doop } (3 BM)     & 48T/24C/2S (Cascade Lake '19; 3)
                  &       64GB & Hadoop cluster used for research \\ \hline
\texttt{cs-igpu } (4 BM) & 64T/32C/2S (Cascade Lake '19; 4)
                  &      512GB & Instructional lab machines with GPUs \\ \hline
\hline  \pagebreak[3]
\texttt{cs-afs1 } (11 BM) &  16T/8C/1S (Cascade Lake '20; 4), 8T/4C/1S (Broadwell '16; 3), 4T/4C/1S (Opteron '10; 2), 20T/10C/1S (Cascade Lake '19), 4T/2C/1S (Clarkdale '10)
                  & 64GB (1), 32GB (4), 16GB (3), 4GB (3) & AFS servers~\cite{openafsfoundationOpenAFS,howardScalePerformanceDistributed1987} \\ \hline
\texttt{cs-afs2 } (6 VM) & 1 vCPU (Broadwell '16; 3), 1 vCPU (Haswell '14; 3)
                  & 2GB (3), 1GB (3) & AFS servers~\cite{openafsfoundationOpenAFS,howardScalePerformanceDistributed1987} \\ \hline
\texttt{cs-sush } (6 VM) & 1 vCPU (Broadwell '16; 5), 4 vCPUs (Broadwell '16)
                  & 16GB (1), 2GB (5) & Backup servers \\ \hline
\texttt{cs-msql } (4 VM) & 1 vCPU (Broadwell '16; 3), 2 vCPUs (Haswell '14)
                  &        2GB  & MySQL databases \\ \hline
\texttt{cs-nfs  } (3~VM,~1~BM) & 1 vCPU (Broadwell '16; 2), 2 vCPUs (Broadwell '16), 48T/24C/2S (EPYC '19)
                  & 128GB (1), 2GB (3) & NFS servers \\ \hline
\texttt{cs-pstg } (2 VM) & 2 vCPUs (Broadwell '16; 2)
                  & 4GB, 2GB    & PostgreSQL servers \\ \hline
\hline  \pagebreak[3]
\texttt{cs-dns  } (6~VM,~1~BM) & 2 vCPUs (Broadwell '16; 3), 1 vCPU (Broadwell '16; 3), 4T/4C/1S (Lynnfield '09)
                  & 8GB (1), 4GB (2), 2GB (3), 1GB (1) & DNS servers, recursers, load balancers \\ \hline
\texttt{cs-mntr } (3 VM) & 4 vCPUs (Broadwell '16; 3)
                  & 16GB (1), 4GB (2) & Network/application monitoring \\ \hline
\texttt{cs-vaul } (3 VM) & 2 vCPUs (Broadwell '16; 3) &         2GB  & HashiCorp Vault secret management servers \\ \hline
\texttt{cs-prx  } (3 VM) & 2 vCPUs (Broadwell '16; 2), 1 vCPU (Broadwell '16) & 4GB (1), 2GB (2)  & Various proxy servers (Haproxy, squid, REST) \\ \hline
\texttt{cs-dhcp } (2 VM) & 2 vCPUs (Broadwell '16; 2) &         2GB  & DHCP hosts \\ \hline
\texttt{cs-conf } (2 VM) & 16 vCPUs (Broadwell '16), 4 vCPUs (Broadwell '16) &   32GB, 8GB  & Configuration management servers \\ \hline
\texttt{cs-vpn  } (1 VM) & 2 vCPUs (Broadwell '16) &         2GB  & VPN server \\ \hline
\texttt{cs-tftp } (1 VM) & 2 vCPUs (Broadwell '16) &         2GB  & TFTPD server used for installs \\ \hline
\texttt{cs-apt  } (1 VM) & 2 vCPUs (Broadwell '16) &         2GB  & Apt-Cacher Debian package cache \\ \hline
\hline  \pagebreak[3]
\texttt{cs-web  } (22 VM) & 2 vCPUs (Broadwell '16; 15), 1 vCPU (Broadwell '16; 4), 2 vCPUs (Haswell '14; 3) & 4GB (2), 2GB (19), 1GB (1)  & Web servers for dept/faculty/staff/student web sites \\ \hline
\texttt{cs-ldap } (7 VM) & 2 vCPUs (Broadwell '16; 3), 4 vCPUs (Broadwell '16; 2), 1 vCPU (Broadwell '16; 2) & 8GB, 4GB (4), 2GB (2)  & LDAP hosts \\ \hline
\texttt{cs-cweb } (5 VM) & 4 vCPUs (Broadwell '16; 3), 2 vCPUs (Broadwell '16; 2) & 16GB, 8GB, 4GB, 2GB (2)  & Containerized web applications \\ \hline
\texttt{cs-mail } (4 VM) & 2 vCPUs (Broadwell '16; 4) & 4GB (3), 2GB (1)  & Mail servers \\ \hline
\texttt{cs-gitl } (3 VM) & 8 vCPUs (Broadwell '16), 4 vCPUs (Broadwell '16), 2 vCPUs (Broadwell '16) & 12GB, 4GB, 2GB  & Gitlab server \\ \hline
\texttt{cs-tckt } (1 VM) & 4 vCPUs (Broadwell '16) &         8GB  & Request Tracker 4 ticket tracker \\ \hline
\texttt{cs-cal  } (1 VM) & 1 vCPU (Broadwell '16) &         2GB  & Department events calendar server \\ \hline
\texttt{cs-rsyn } (1 VM) & 2 vCPUs (Broadwell '16) &         2GB  & rsync server for external collaborators \\ \hline
\caption{Description and count of instrumented nodes. \codesm{cs-*} are department infrastructure hosts; \codesm{hpc-*} are high-performance computing infrastructure. ``BM/VM'' indicates whether the node is bare-metal or a virtual machine. \label{fig:nodes}}
\end{longtable}
}

\twocolumn

\end{document}